\documentclass[10pt, conferenc, final]{elsarticle}
\usepackage{xspace}
\usepackage{booktabs}
\usepackage{graphicx}
\usepackage{amsmath}
\usepackage{amssymb}
\usepackage{color}
\usepackage{ifpdf}
\usepackage{float}
\usepackage[utf8]{inputenc}
\usepackage{multirow}
\usepackage{rotating}
\usepackage{array}
\usepackage{mathtools}

\usepackage{moresize}
\usepackage{fancyvrb}
\usepackage{url}
\usepackage{booktabs}
\usepackage{listings}   
\usepackage{paralist}    
\usepackage{wrapfig}    
\usepackage{multirow}
\usepackage{ifpdf}
\usepackage{srcltx}
\usepackage{xspace}
\usepackage{keyval}  
\usepackage{color}
\usepackage{soul}
\usepackage{caption}
\usepackage{subcaption}
\usepackage[export]{adjustbox}

\definecolor{listinggray}{gray}{0.95}
\definecolor{darkgray}{gray}{0.7}
\definecolor{commentgreen}{rgb}{0, 0.4, 0}
\definecolor{darkblue}{rgb}{0, 0, 0.6}
\definecolor{purple}{rgb}{0.6, 0, 0.6}
\definecolor{middleblue}{rgb}{0, 0, 0.75}
\definecolor{darkred}{rgb}{0.4, 0, 0}
\definecolor{brown}{rgb}{0.5, 0.5, 0}
\definecolor{dkgreen}{rgb}{0,0.5,0}
\definecolor{orange}{rgb}{1,.5,0}
\definecolor{dandelion}{cmyk}{0,0.29,0.84,0}

\usepackage[normalem]{ulem}
\makeatletter
\def\cyanuwave{\bgroup \markoverwith{\lower3.5\p@\hbox{\sixly \textcolor{cyan}{\char58}}}\ULon}
\def\reduwave{\bgroup \markoverwith{\lower3.5\p@\hbox{\sixly \textcolor{red}{\char58}}}\ULon}
\def\blueuwave{\bgroup \markoverwith{\lower3.5\p@\hbox{\sixly \textcolor{blue}{\char58}}}\ULon}
\font\sixly=lasy6 
\makeatother

\ifpdf
 \DeclareGraphicsExtensions{.pdf, .jpg, .tif, .png}
\else
 \DeclareGraphicsExtensions{.png, .ps}
\fi

\newcommand{\B}[1]{\textbf{#1}\xspace}
\newcommand{\I}[1]{\textit{#1}\xspace}
\newcommand{\T}[1]{\texttt{#1}\xspace}

\newcommand{\UP}{\vspace*{-1em}}

\newif\ifdraft
\drafttrue
\draftfalse
\ifdraft
\usepackage{xcolor}
\newcommand{\jhanote} [1]{{\textcolor{red}   { ***SJ\@:  #1 }}}
\newcommand{\amnote}  [1]{{\textcolor{blue}  { ***AM\@:  #1 }}}
\newcommand{\mtnote}  [1]{{\textcolor{orange}{ ***MT\@:  #1 }}}
\newcommand{\mingnote}[1]{{\textcolor{purple}{ ***MTH\@: #1 }}}
\newcommand{\note}    [1]{{\textcolor{green} { ***Note:  #1 }}}
\else
\newcommand{\onote}[1]{}
\newcommand{\terminology}[1]{}
\newcommand{\jhanote}[1]{}
\newcommand{\amnote}[1]{}
\newcommand{\katznote}[1]{}
\newcommand{\mtnote}[1]{}
\newcommand{\mingnote}[1]{}
\newcommand{\note}[1]{}
\fi

\newcommand{\up}{\vspace*{-0.5em}}

\newcommand{\synapse}{Synapse\xspace}
\newcommand{\Synapse}{Synapse\xspace}

\newcommand{\rp}{RADICAL-Pilot\xspace}
\newcommand{\enmd}{Ensemble Toolkit\xspace}

\newcommand{\titan}{\I{Titan}}
\newcommand{\comet}{\I{Comet}}
\newcommand{\stampede}{\I{Stampede}}
\newcommand{\supermic}{\I{Supermic}}
\newcommand{\archer}{\I{Archer}}
\newcommand{\prof}{\I{Thinkie}}

\DefineShortVerb{\|}

\DefineVerbatimEnvironment{mycode}{Verbatim}
{
  label=Code Example,
  fontsize=\small,
  frame=single,
  framerule=1pt,
  framesep=1em,
  numbers=left,
  gobble=2
}

\DefineVerbatimEnvironment{myio}{Verbatim}
{
  fontsize=\small,
  frame=lines,
  framerule=1pt,
  framesep=0.5em
}

\newenvironment{shortlist}{
  \begin{itemize}
}{
  \end{itemize}
}

\title{Synapse: Synthetic Application Profiler and Emulator} 
 
\author[add2]{Andre Merzky  } 
\author[add2]{Ming Tai Ha   } 
\author[add2]{Matteo Turilli} 
\author[add2,add3]{Shantenu Jha  } 

\address[add2]{%
    RADICAL Laboratory, 
    Electric and Computer Engineering\\
    Rutgers University, 
    New Brunswick, NJ, USA\\
}

\address[add3]{%
    Center for Data Driven Discovery, 
    Computational Science Initiative\\
    Brookhaven National Laboratory, NY, USA\\
}

\begin{document}

%

\begin{abstract}

Motivated by the need to emulate workload execution characteristics on 
high-performance and distributed heterogeneous resources, we introduce  
\synapse. \synapse is used as a proxy application (or "representative
application") for real workloads, with the advantage that it can be
tuned in different ways and dimensions, and also at levels of granularity that
are not possible with real applications. \synapse has a platform-independent 
application profiler, and has the ability to emulate profiled workloads 
on a variety of resources. Experiments show that the automated profiling 
performed using \synapse captures an application's characteristics with
high fidelity. The emulation of an application using \synapse can 
reproduce the application's execution behavior in the original runtime 
environment, and can also reproduce those behaviors on different run-time 
environments.


\end{abstract}


\maketitle

%
\section{Introduction}\label{sec:intro}
A large body of research in high-performance and distributed computing is
concerned with the design, implementation and optimization of tools, runtime
systems and services in support of scientific applications. These tools,
systems and services, often subsumed under the term ``middleware'', require
extensive and continuous development, testing and optimization based on
various, real-life applications and production-grade infrastructures.

The use of scientific applications for developing middleware present three
main challenges: large amount of deployment requirements; domain-specific
knowledge for successful and representative execution; and limited
portability and scalability. Synthetic applications---also known as
Skeletons, Representative, or Artificial Applications---help to address these
challenges by working as proxies of real-life applications.

Synthetic applications capture the relevant properties of applications,
minimizing deployment requirements and eliminating the need of
domain-specific knowledge for their execution. Further, they enable tuning
application parameters relevant to scientific middleware development. Often,
parameters like precise runtime, number of computational cycles, memory
footprint or I/O patterns cannot be precisely tuned in real-life
applications.

A tradeoff in the design and implementation of synthetic applications for use
as proxy applications is the need to be simple and general-purpose and to be
able to emulate the behavior of each application with the highest level of
accuracy and fidelity possible. Achieving this level of accuracy and
fidelity is particularly challenging when an emulation is used on multiple
heterogeneous resources. It is even more challenging when the resources used
for emulation are different from the resource on which the application(s) was
profiled.

In response to these requirements and tradeoffs, we have developed
\synapse{}: a \I{SYNthetic Application Profiler and Emulator}. \synapse{} is
primarily motivated by the need for automated and system-independent
application profiling in computational science, where the multitude and
generality of applications and platforms are more important than cycle-level
accuracy and fidelity. Accordingly, \synapse{} is designed to provide uniform
profiling capabilities across a range of application types, tools and
services, while achieving a sufficient level of accuracy and fidelity.

\synapse{} acts as a proxy application to circumvent the limitations
and complexity of scientific applications. For example, scientific
applications are not infinitely malleable because of the fixed and 
often discrete physical sizes of input systems; they have limited 
tunability as parameters can be modified only in discrete steps over a 
limited range of values. \synapse{} can profile an application for given 
parameter values \textit{and} can emulate the execution behavior of
the same application for different parameter values.

\synapse{} is designed to ``profile once, emulate anywhere'': \synapse{}
determines the application's resource consumption by running a sample-based,
black-box profiler of the application on a machine with specific tools and
permissions, and replays the observed resource consumption patterns on an
arbitrary machine. In emulation mode, \synapse{} attempts to consume the same
amount of system resources (clock cycles, memory reads/writes, packet
sent/received) as the original application, while preserving the overall
pattern, granularity and sequence of the respective operations.

\synapse{} provides basic kernels functions to emulate how different system
resources are consumed. While these kernels can be controlled to consume the
specified amount of system resources precisely, they are less precise on how
those resources are consumed, i.e., in what chunkiness, granularity and
order. \synapse{} provides the ability for users to write their own kernels
to control more tightly how system resources are consumed. Thus, even when
different applications consume the same amount of some system resource,
\synapse{} can be tuned to consume system resources in a way that better
reflects how a specific application consumes the same resource.


This paper presents the design of \synapse{} and progress towards an
implementation that is robust and usable. Further, we presents 
six experiments to validate that \synapse{}'s automated profiling indeed
captures the application characteristics with fidelity. The experiments also
show that \synapse{}'s emulation reproduces the application characteristics
in the original runtime environment, as well as on different resources and
runtime systems.

While \synapse, and in particular its profiler, is not designed to achieve
the same accuracy as other established approaches (e.g.,\,
Vtune~\cite{reinders2005vtune}), our experiments support the claim that
\synapse{}'s emulation has sufficient fidelity, generality and tunability to
make it a useful instrument for the development of middleware to support
computational science research.

In Section~\ref{sec:overview} we outline three application and systems use cases
that have motivated the development of \synapse{}. In Section~\ref{sec:arch}, we
discuss the design and architecture of \synapse{}, followed by a discussion of
selected implementation details in Section~\ref{sec:impl}. Experiments are
discussed in Section~\ref{sec:exp}, followed by future and related work.

%
\section{A Case for System-Independent Profiling and
Emulation}\label{sec:overview}
The development of tools for computational science and large-scale computer
science experiments needs proxy applications that provide flexible and
tunable capabilities as well as being portable across resource types. We
outline three 
use cases for proxy applications, each highlighting a different requirement.

\subsection{High-Performance Task-Parallel Computing} 

Traditionally, high-performance computing (HPC) systems have been optimized
to support mostly monolithic workloads. The workload of many scientific
applications however, are comprised of spatially and temporally heterogeneous
tasks that are often dynamically inter-related~\cite{note4}. These workloads
can benefit from being executed at scale on HPC resources, but a tension
exists between their resource requirements and the capabilities of HPC
systems and their usage policies. We addressed this tension by developing
RADICAL-Pilot~\cite{review_radicalpilot_2015}, a scalable and
interoperable pilot system~\cite{pilot-review}.

\rp provides a runtime system designed to support a large number of
concurrent tasks with low start-up overhead. \rp is agnostic to the specific
properties of the executed tasks, supporting many-node parallelism as well as
single core tasks, and both short and long running tasks. Many components of
\rp are designed and parameterized to provide balanced performance while
being as agnostic as possible to task and resource properties. For example,
\rp's task execution component, the RP Agent, has to be engineered for
optimal resource utilization while maintaining full generality in many
different dimensions, like MPI/non-MPI\@;
OpenMP/multi-threaded/single-threaded; CPU/GPU/accelerators;
single-node/multi-node; ho\-mo\-ge\-neous/het\-ero\-ge\-neous tasks and
clusters; or different batch systems.



We can support the design and testing of \rp by tuning the properties of a
single proxy application instead of refactoring multiple scientific
applications. For example, we can profile a single-threaded scientific
application and then emulate it as a mixed OpenMPI or MPI proxy application
on an arbitrary resource. Analogously, we can increase the amount of memory
required by the same proxy application to a specific value, even if the
science problem of the profiled application does not require that amount of
memory.

\subsection{Abstractions and Middleware for Distributed Computing} 

In spite of significant progress in scientific distributed computing over the
past decade, there are few general-purpose abstractions that support a
principled development of middleware for large-scale and distributed
execution of applications. As a consequence, many software point-solutions
exist that are tailored to specific workloads or resource types but few
middleware are capable of supporting arbitrary applications on arbitrary
types of resources. The DOE AIMES project contributed to address this problem
by defining general-purpose abstractions and developing a middleware for
distributing the execution of large-scale applications across multiple
resources~\cite{ipdps-2016}.




AIMES assumed the use of third-party tools to manage dependences among tasks
of scientific application. As a consequence, the main challenges in
generalizing the capabilities of the AIMES middleware to different
applications on different types of resource were primarily related to
implementation and deployment. A proxy application that can emulate the
execution behavior of actual workloads would play an important role in the
validation and extension of base AIMES abstractions and middleware. Proxy
applications have the advantage of capturing relevant application properties
without exposing the complexity of running these applications on distinct
platforms.

\subsection{Toolkits for Computational Science}

Many scientific applications in the field of molecular sciences,
computational biology~\cite{preto2014fast},
astrophysics~\cite{sirko2005initial}, weather
forecasting~\cite{bauer2015quiet}, and
bioinformatics~\cite{martin2010rnnotator} are increasingly reliant on
ensemble-based methods to make scientific progress. Ensemble-based
applications vary in the degree of coupling and dependency between tasks, and
in heterogeneity across tasks. In spite of the apparent simplicity of running
ensemble-based applications, the scalable and flexible execution of a large
set of tasks is non-trivial. 

\mtnote{Add a paragraph to describe EnTK?}

As a consequence of complexity and many degrees-of-freedom, the challenges
and the growing importance and pervasiveness of ensemble applications, we
designed and implemented \enmd~\cite{enmd}. Similar to the previous two
use-cases, a proxy application would provide a lightweight and highly tunable
workload so as to simplify and design \enmd for general purpose workloads. In
addition, a proxy application would provide the ability to vary the duration
and number of task instances between different stages of the application and
change the coupling between tasks; this is an important characteristics of
applications used for advanced sampling~\cite{enmd_docs}.

\mtnote{I am worried that the three cases we use to motivate proxy
applications are all from our own research.}
\amnote{Yeah, that is true - but OTOH we don't have any synapse users, and I
at least have not seen many synthetic applications in the wirld, yet... :/}

%
\section{\Synapse Scope and Architecture}\label{sec:arch}
A finer-grained analysis of the aforementioned use cases, results in the
following requirements on the profiling and emulation stages of \synapse.

\subsection{Requirements on Application Profiling}\label{sec:prof_req}

We state four requirements for correct profiling:

\begin{shortlist}

  \item \B{P.1 Minimal Self-Interference:} Profiling does not influence the
      resulting profile of an executable;

  \item \B{P.2 Low Overhead:} Profiling does not influence the runtime behavior
      of the profiled executable;

  \item \B{P.3 Black-box Approach:} Profiling does not require any
      changes in application code, and minimal, non-intrusive changes in
      application runtime environment;

  \item \B{P.4 Consistency:} Repeated profiling of the same application, in
      the same environment, yields consistent profiles;

  \item \B{P.5 Reproducibility:} Profiles are usable to reproduce (emulate)
      the application's runtime behavior.
        
\end{shortlist}


\subsection{Requirements on Application Emulation}\label{sec:emu_req}

Requirements for application emulation are:

\begin{shortlist}

  \item \B{E.1 Fidelity:} Application emulation must exhibit the same
      essential runtime characteristics as the execution of the actual
      application. Amongst others, we specifically expect emulation execution
      time \(T_x\) to correspond to the application \(T_x\).

  \item \B{E.2 Portability:} An application can be emulated on resources
      other than the one used for its profiling.

  \item \B{E.3 Malleability} Application emulation can be tuned along a
      variety of dimensions, specifically including those not covered by the
      profiling or by the original application.

\end{shortlist}

\subsection{Synapse Architecture}\label{ssec:arch}

\Synapse is a research prototype used to support other research projects.  As
such, it is subject to frequent changes in target use-cases, requested
features and systems to be supported. We chose an architecture that is
modular for both the profiling and the emulation part of \synapse to support
dynamic development.

\begin{figure}[ht]
 \centering
 \includegraphics[width=0.85\textwidth]{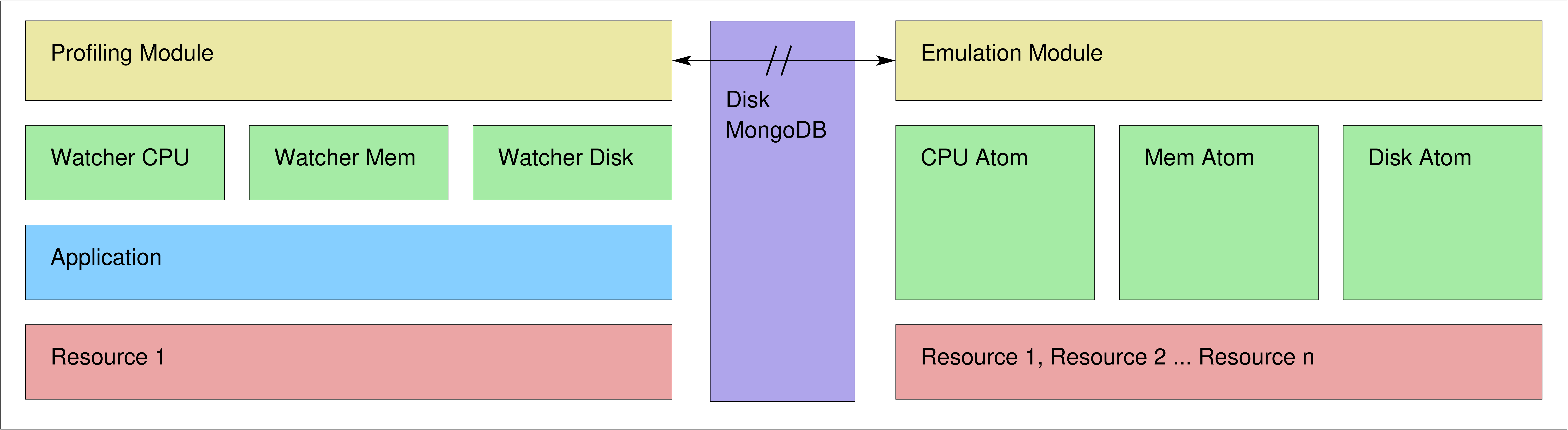}
 \caption{\textbf{\Synapse Architecture:} The profiling module manages 
     `Watcher plugins' which observe the runtime behavior of an
     application execution.  The emulation module interprets the resulting
     profiles to control `Emulation Atoms' which consume resources
     accordingly.  Profiles are stored on and read from persisten storage.
 }\label{fig:arch}
\end{figure}

Figure~\ref{fig:arch} illustrate the architecture of \synapse. The modularity
of the \synapse profiling is provided by extensible and exchangeable
\I{Watcher} plugins, each profiling the use of a specific type of system
resource. The counterpart in the \synapse emulation module are pluggable
emulation \I{Atoms} which can emulate the consumption of the same types of
system resource types as profiled by the Watchers. 


%
\section{Implementation of \Synapse}\label{sec:impl}
\synapse is mainly written 
in Python 
with some parts 
written in C and assembly. 
Implemented as a Python module, 
in its basic usage mode \synapse provides two methods:

\begin{myio}
radical.synapse.profile(command, tags=None)
radical.synapse.emulate(command, tags=None)
\end{myio}

\noindent where |command| is either a shell command line or a Python
callable. \Synapse also provides a set of command line tools which are
wrappers around certain configurations and combinations of the |profile| and
|emulate| methods.

The |profile| method profiles an application executable and stores the results
on disk or in a MongoDB database.  The application startup command and custom
tags are used as search index for that database\footnote{Tags are used to
distinguish profiles where the application has been executed by the same
command line, but where configuration files or other variables change the
application's workload.}.  When collecting multiple profiles for the same set
command and tag combination, \synapse can perform some basic statistics analysis
on the resource consumption recorded across those profiles.

\mtnote{This sentence is very compressed, I read it three times and I still
struggle to understand what it wants to say. What is `different semantic
content' when referred to files in which (I guess) we store
instructions/directives/variables for application control?}
\amnote{better?}\mtnote{Better but I think we may want to iterate further if
given the opportunity.}\amnote{better? :-)}
 
The |emulate| method utilizes emulation \I{atoms} (see
Section~\ref{ssec:arch}), each being a fine-grained and tunable software
element that consumes one type of system resource. \Synapse uses the
command/tag combination specified in the |emulate| call to search the
database for a matching profile. Once a profile is found, \synapse retrieves
the profile and feeds all samples it contains to the emulation atoms in the
order in which the samples have been collected. The sample ordering is an
essential element of the fidelity of the emulation (see
Section~\ref{sec:sampling}),


\subsection{Implementation of \Synapse Profiling}
 
The \synapse profiler relies on several system utilities. Primarily, \synapse
uses the |perf-stat| utility to inspect CPU activity, the |/proc/| filesystem
to read system counters on memory and disk I/O, and the POSIX |rusage| call
to obtain runtime process information. The different information providers
are implemented as plugins, \synapse is thus extensible with additional
profiling metrics (see discussion of future work in
Section~\ref{sec:future}). Those plugins are structured as follows:

 \begin{myio}
 class WatcherClass(WatcherBase):
    def __init__    (self, pid):    ...
    def pre_process (self, config): ...
    def sample      (self):         ...
    def post_process(self):         ...
    def finalize    (self):         ...
 \end{myio}

|pre_process| and |post_process| set up and tear down any profiling
environment for that watcher.  The |sample| method is invoked at regular
intervals by the main \synapse profiling loop.  In the |finalize| method, the
plugin can access raw profiling results of other watchers, in order to
perform some further post-processing\mtnote{post-processing?}\amnote{yes}.
While this can create some dependencies between plugins, it prevents the
duplication of measurements, such as overall runtime. Each watcher plugin
runs in its own thread:

\begin{myio}
def run(self):
    self.pre_process(self._config)

    while not self._terminate.is_set():
        now = timestamp()
        self.sample(now)
        time.sleep(1.0/self._sample_rate)

    self.post_process()
\end{myio}

Once \synapse spawns the application process, it communicates the application
process' PID to the watcher threads, which monitor the application process.
There is a small delay between process spawning and start of profiling but
the process itself is wrapped into the POSIX tool |time -v|, which allows us
to correct some of the effects of that offset. Note that 
other effects are found to be too small to matter: The first watcher sample
is usually collected at around $0.005$ seconds after startup.

The sample rate is globally controlled via a configuration option and it is
uniform over all watchers. \synapse can at most gather one sample every $100
ms$ (i.e.,~10 samples per second), which coincides with the sampling limit of
|perf stat|; there is no lower bound to the sampling rate.
Section~\ref{sec:exp} discusses the impact of different sampling rates on
profiling overhead, profiling accuracy, and emulation fidelity.

Profile data are collected as time series. The timestamps of the different
watchers are not synchronized, and can drift relative to each other over
time.  We found this preferable to an increased profiling overhead due to
synchronization.  The individual time series are combined during
postprocessing and pushed into a MongoDB, or 
written to disk.

\subsection{Implementation of \Synapse Emulation}

At its core, the \synapse emulation framework consists of a set of emulation
\I{atoms}, where each atom uses a C or assembly function (kernel) to consume
a specific type of system resource. Currently, |compute|, |memory|, |storage|
and |network| atoms have been implemented. The \synapse emulation atoms are
driven by a global loop which feeds sequences of profile samples to the atoms
for emulation.  The sample granularity is the same as used for profiling: the
profiling sampling rate thus not only determines the accuracy of the
profiling itself, but also influences the emulation fidelity.  Each atom runs
in as a separate thread to enable the concurrent utilization of resources
(in the limited way threading is supported by Python).  Atom implementations
are interchangeable.


The default |compute| atom implementation contains a kernel running a loop of
assembly code that performs a matrix multiplication with small matrices (they
fit into the CPU cache) very efficiently.  The loop's efficiency represents
the maximum efficiency at which this atom can emulate.

Other kernels for |compute| atoms are implemented in C, and perform matrix
multiplications on data which do not usually fit into the CPU caches.  Those
kernels have a lower efficiency, but they represent actual application codes
more realistically. Users can provide additional |compute| kernels, coded in
Python, C or Assembly, which grants full flexibility and control over how
data are accessed when performing computation. Experiments in
Section~\ref{sec:exp} show that this modularity enables Synapse to emulate
applications with greater fidelity.



%
%

The default |memory| and |storage| atoms perform the canonical |libc|
operations |malloc|, |free|, |read| and |write|.  Those operations use block
sizes that can be tuned but, at the moment, those block sizes are not related
to the recorded profiles. This introduces discrepancies compared to the
emulated application because system performance depends on the block size of
I/O operations.  Our current assumption is that application codes are
generally aware of this, and attempt to use large block sizes where possible,
and that small reads/writes are dominantly served by disk caches, and have
thus relatively small impact on the overall performance.

We acknowledge that this naive assumption is likely to break (to a varying
degree) for certain types of applications whose performance is bound by
specific I/O patterns.  To mitigate this effect, and also to further the goal
of malleability, users can configure synapse to use specific I/O block
sizes, Experiment E.5 in Section~\ref{sec:exp} investigates the tunabililty
of I/O blocks in more detail.
The \synapse profiler features an experimental watcher plugin that can, in
principle, infer block sizes of disk I/O operations using |blktrace|.  We
consider using this data in \synapse emulation when applications require that
granularity to be future work (see Section~\ref{sec:future}).  Similarly,
memory I/O granularity can significantly influence application performance,
and \synapse will have to evolve to accommodate applications sensitive to
it.

\mtnote{Not iterated because partial. Ready to iterate further when
completed.}\amnote{please do.}\mtnote{Done.}

\subsection{Profiling Metrics}\label{sec:metrics}

\begin{table}
\begin{center}
\scalebox{0.75}{
\begin{tabular}{llcccc}
  \toprule
    \B{Resource} & \B{Metric}              & \B{Tot.} & \B{Samp.} & \B{Der.} & \B{Emul.}\\\midrule
    \B{System }  & number of cores         &  \T{ + } &   \T{ - } &  \T{ - } &  \T{ - } \\
                 & max CPU frequency       &  \T{ + } &   \T{ - } &  \T{ - } &  \T{ - } \\
                 & total memory            &  \T{ + } &   \T{ - } &  \T{ - } &  \T{ - } \\
                 & runtime                 &  \T{ + } &   \T{ + } &  \T{ - } &  \T{ - } \\
                 & system load (CPU)       &  \T{ + } &   \T{ - } &  \T{ - } &  \T{ + } \\
                 & system load (disk)      &  \T{ - } &   \T{ - } &  \T{ - } &  \T{ + } \\
                 & system load (memory)    &  \T{ - } &   \T{ - } &  \T{ - } &  \T{ + } \\\midrule
    \B{Compute}  & CPU instructions        &  \T{ + } &   \T{ + } &  \T{ - } &  \T{ + } \\
                 & cycles used             &  \T{ + } &   \T{ + } &  \T{ - } &  \T{ + } \\
                 & cycles stalled backend  &  \T{ + } &   \T{ + } &  \T{ - } &  \T{ - } \\
                 & cycles stalled frontend &  \T{ + } &   \T{ + } &  \T{ - } &  \T{ - } \\
                 & efficiency              &  \T{ + } &   \T{ + } &  \T{ + } &  \T{(+)} \\
                 & utilization             &  \T{ + } &   \T{ + } &  \T{ + } &  \T{ - } \\
                 & FLOPs                   &  \T{ + } &   \T{ + } &  \T{ + } &  \T{ + } \\
                 & FLOP/s                  &  \T{ + } &   \T{ + } &  \T{ + } &  \T{ - } \\
                 & number of threads       &  \T{ + } &   \T{ - } &  \T{ - } &  \T{(+)} \\
                 & OpenMP                  &  \T{(+)} &   \T{ - } &  \T{ - } &  \T{ + } \\\midrule
    \B{Storage}  & bytes read              &  \T{ + } &   \T{ + } &  \T{ - } &  \T{ + } \\
                 & bytes written           &  \T{ + } &   \T{ + } &  \T{ - } &  \T{ + } \\
                 & block size read         &  \T{ - } &   \T{(+)} &  \T{ - } &  \T{ + } \\
                 & block size write        &  \T{ - } &   \T{(+)} &  \T{ - } &  \T{ + } \\
                 & used file system        &  \T{ + } &   \T{ - } &  \T{ - } &  \T{ + } \\\midrule
    \B{Memory}   & bytes peak              &  \T{ + } &   \T{ + } &  \T{ - } &  \T{ - } \\
                 & bytes resident size     &  \T{ + } &   \T{ + } &  \T{ - } &  \T{ - } \\
                 & bytes allocated         &  \T{ + } &   \T{ + } &  \T{ + } &  \T{ + } \\
                 & bytes freed             &  \T{ + } &   \T{ + } &  \T{ + } &  \T{ + } \\
                 & block size alloc        &  \T{ - } &   \T{(-)} &  \T{ - } &  \T{(-)} \\
                 & block size free         &  \T{ - } &   \T{(-)} &  \T{ - } &  \T{(-)} \\\midrule
    \B{Network}  & connection endpoint     &  \T{(-)} &   \T{(-)} &  \T{ - } &  \T{(+)} \\
                 & bytes read              &  \T{(-)} &   \T{(-)} &  \T{ - } &  \T{(+)} \\
                 & bytes written           &  \T{(-)} &   \T{(-)} &  \T{ - } &  \T{(+)} \\
                 & block size read         &  \T{ - } &   \T{(-)} &  \T{ - } &  \T{(-)} \\
                 & block size write        &  \T{ - } &   \T{(-)} &  \T{ - } &  \T{(-)} \\\midrule
\end{tabular}}
\caption{
    \B{List of \synapse metrics and their usage}\newline
    \B{Sampl.:} sampled over time;
    \B{Der.:} derived from other metrics;
    \B{Tot.:} integrated total over runtime;
    \B{Emul.:} used in emulation;
    \B{\T{(+)}:} partial;
    \B{\T{(-)}:} planned.
 }
\UP\label{tab:metrics}
\end{center}
\end{table}


Table~\ref{tab:metrics} shows the metrics measured by \synapse for the three
types of resources that are currently profiled: compute (CPU), storage
(disk), and memory. Additionally, \synapse records several types of system
information, such as number and type of CPU cores, available memory, system
load, etc.  Some of those metrics are used to compute derived metrics, for
example, the CPU type and clock speed determines the maximum number of
operations per second, which yields CPU utilization and efficiency when
combined with the observed number of used and stalled instructions cycles.

\synapse is able to force an artificial CPU, disk and memory load onto the
system while emulating an application, thus emulating the application
execution in a stressed environment (similar to the Linux utility
\T{'stress'}). Currently, we do not measure the original disk and memory
load 
on the system, so these 
factors have to be specified manually, and are currently used to confirm
\synapse's viability on stressed systems. Artificial loads have not been used
in the experiments presented in this paper, and are thus not further
discussed. 

Several metrics, such as CPU efficiency and utilization, are marked as
\I{derived}: they are not directly reported by the system, but calculated
from other, primary metrics.  We use the following formula to compute CPU
efficiency:\\[-2em]

\begin{eqnarray*} \mathit{efficiency} &=& cycles_{used} /
    cycles_{spent}\\ &=& cycles_{used} / (cycles_{used} +
    cycles_{wasted}) \end{eqnarray*}\\[-2em]

\noindent We interpret the value that |perf stat| reports as |cycles| as as
$cycles_{used}$, and we consider all stalled cycles
($cycles\_stalled\_frontend~+~cycles\_stalled\_backend$) as wasted
($cycles_{wasted}$),
indicating that cycles are counted toward the application execution, but did not
contribute to its progression.  Note that this definition can potentially count
some cycles twice, as the CPU frontend and backend (as defined by \T{perf}) run
concurrently in the same CPU cycle, and both can stall at the same time. Also,
stalled cycles can overlap with used cycles as the backend can be busy while the
frontend stalls.  Accordingly, this metric 
considers used cycles to contribute to higher efficiency, and any stalling to
lower efficiency.

Similarly, we compute CPU utilization as:\\[-2.5em]

\begin{eqnarray*}
   utilization &=& cycles_{used} / cycles_{max}
\end{eqnarray*}\\[-2.5em]

\noindent where $cycles_{max}$ is derived from the maximum possible number of
cycles, as determined by CPU architecture and clock speed.  \Synapse does not
sample the CPU clock speed (modern CPUs can adapt clock speed to load to
preserve energy), and we do not take any background CPU activity (by the
system or other applications) into account.  The derived utilization is still
a useful metric as it exposes the expected monotonic behavior toward
faster/slower execution, but it is not necessarily comparable to similar
metrics derived by other system utilities like |ksysguard| and the like.

Table~\ref{tab:metrics} includes several metrics that are currently planned
or only partially implemented.  Specifically, it lists network interactions,
which \synapse can emulate to some extent 
but which are not yet meaningfully profiled. CPU efficiency is listed as
`partially supported' for emulation: \synapse is able to tune the CPU load
toward a certain efficiency value, but that tuning is currently manually set
(experiments presented in the paper use default values for all
tunable settings).

\subsection{The Effects of Sampling}\label{sec:sampling}

\begin{figure}
 \centering
 \includegraphics[width=0.85\textwidth]{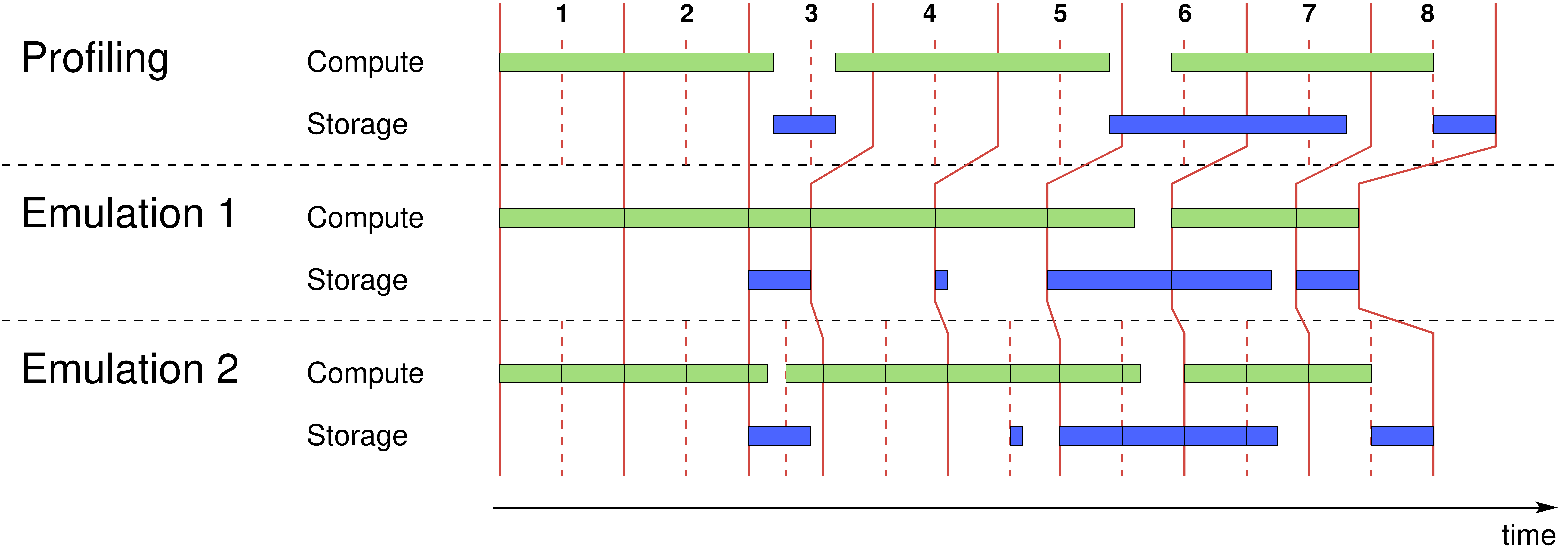}
 \caption{\textbf{Sampling Effects:}
   the profiling samples (top) show that the application used a mix of serial
   and concurrent CPU (green) and disk (blue) operations.  Solid red lines
   represent profiling sample boundaries; broken red lines represent sample
   boundaries at doubled sampling rate.  Emulations 1 and 2 (middle, bottom)
   replay different sample types (compute, storage) concurrently, thus
   removing some of the serialization of the original resource consumption
   (see sample numbers 3, 8).  Emulation 2 replays the higher sampling rate,
   thereby reducing that effect by partially re-introducing the serialization
   of the original resource consumption (see again samples 3,
   8).\label{fig:sampling_1} }
\end{figure}

The effects of sampling are illustrated in Figure~\ref{fig:sampling_1}.
Profiling metrics are gathered at roughly equidistant points in time, for
different types of resources.  Emulation follows the same clustering, but
\I{disregards all timing information}. This is consistent with the purpose of
emulation: consuming the same amount of resources as the emulated
application, not reproducing the same timings.
We will discuss several detail of this figure below.

Figure~\ref{fig:sampling_1} illustrates that profiled resource consumptions
may or may not fill a complete sampling period. When a resource type fills a
sampling period, one can expect that the application performance is dominated
by the interactions with that resource type for that sample (e.g., Compute
for sample 3, Storage for sample 6). In the general case, a sampling period
will capture several full or partial resource consumption types which may or
may not occur concurrently.

What resource consumption operation is accounted for in what sample depends
on several parameters: applications often employ techniques to hide I/O
latency, such as caching or asynchronous operations, and the operating system
itself uses latency hiding (caches, read-ahead, branch prediction etc).  In
those cases, actual system activity can occur before or after the application
code requests it.

During \synapse's emulation, all resource consumptions for a specific sample
are started immediately and concurrently upon starting that sample, without
any ordering in between resource types.  Emulation samples end when the last
resource consumptions is completed for that sample, and then the emulation
for the next sample is started (see samples 3 \& 4 in
Figure~\ref{fig:sampling_1}).  Resource consumptions that are not concurrent
in the application \I{are} concurrent in the emulation (see samples 3 and 8),
thus yielding potential emulation speedup. Smaller sampling intervals reduce
that effect (see Emulation 2 in the figure, alternative samples 3 \& 8).
  

In many cases, one type of resource consumption is a semantic requirement for
another type.  For example, an application needs to read data from a disk
before being able to compute on those data; 
allocate memory before reading data from disk into that memory; 
perform computation before being able to write results to a disk; etc. The
code-agnostic sampling approach used by \synapse does not allow to directly
detect such dependencies. However, parts of those dependency information are
implicitly captured: Operations observed in a sample at time $t_n$ can only
depend on resource consumption at samples from $t_{n-1}$ or earlier. By
ensuring that the emulation respects sampling order across resource types,
\synapse will honor the dependencies thus captured, by preserving sampling
order across resource types.

\begin{figure}
  \centering
  \includegraphics[width=0.85\textwidth]{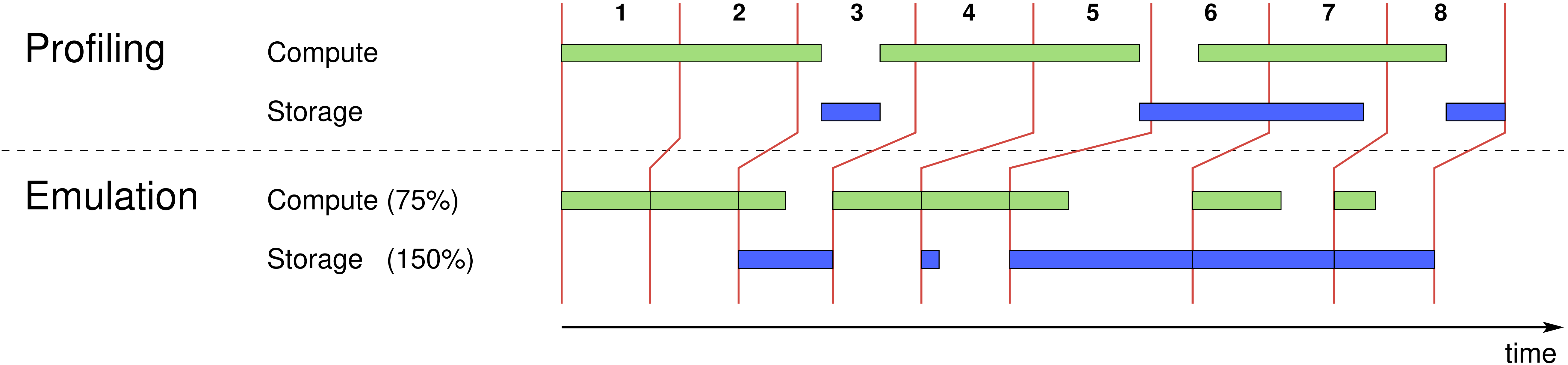}
  \caption{\textbf{Sample Portability:} the same profile samples as
      in figure~\ref{fig:sampling_1} is shown on top.  The bottom
      shows the emulation on a resource with different performance
      (CPU is $25\%$ faster, disk is $50\%$ slower).  The dominating
      resource type switches for some samples (3, 6--8), but the
      overall order of operations is preserved.\label{fig:sampling_2} 
} 
\end{figure}

\synapse profile samples are designed to be portable, i.e., they can
be used to emulate the application on resources other than the profiling
resource.  Figure~\ref{fig:sampling_2} illustrates that the implicit
dependencies captured in the sampling order preserves the order of the
original application activities even on machines with very different
performance characteristics, reflecting that the dominating contribution to
the application\'s \(T_x\) can differ per machine.  In the figure, we see
that the emulation is performed on a machine with faster CPU, but slower
disk.  While sample 7 on the originally profile machine is, for example,
dominated by the application's CPU utilization, the same sample is storage
I/O dominated during emulation.

\subsection{Scope and Limitations}\label{sec:limits}

Sections~\ref{sec:intro} and~\ref{sec:overview} motivated the scope for which
\synapse was defined.  This subsection makes this scope more specific; we
list the set of conditions under which \synapse is expected to operate, or
under which it is not.
 
\paragraph{\B{Resource Limits:}} \synapse does not impose any specific limit
on the amount of resource consumptions it can emulate. Obviously, memory
emulation is limited by the amount of available system memory, disk I/O is
limited by the amount of available storage, etc.

\paragraph{\B{Application Semantics:}} \synapse watches
\I{application behavior}---it explicitly does not inspect the application at
the code or system call level, and thus has no knowledge whatsoever of
application semantics.  This limits the applicability of \synapse's profiling
abilities in some contexts. For example, the POSIX system call |sleep(3)|
will consume a very small number of flops (or cycles), but will show
significant contributions to \(T_x\). An inspection on different layers
(code, libc call, OS signals etc.) could reveal that behavior, but that is
considered out of scope for \synapse 
as it cannot reproduce those operations.

This shortcoming is partially mitigated during emulation, in that users can
select application specific kernels.  By selecting 
a kernel 
the user controls type, amounts and order in which compute (and other)
operations are performed when emulating an application.  For example, a user
could provide an emulation kernel which performs |sleep(n)| or some
equivalent operation. Kernel selection is not automated though, and requires
manual configuration by the user, which implies 
knowledge about the details of profiling and emulation.

\paragraph{\B{Application Granularity:}} An important side effect of
external, sampled measurements is that application activities are not
resolved beyond a certain granularity.  For example, \synapse can measure the
number of bytes written to disk in a certain period of time, but actual I/O
performance can vary significantly depending on \I{how exactly} those I/O
operations are executed: a large number of small, scattered I/O operations
will often be much slower than a small number of large I/O operations. 
\Synapse \I{emulation} allows to adjust block sizes for I/O:  All
experiments in this paper were performed with the default (static) settings,
apart from experiment E.5 which specifically investigates the influence of
the I/O block sizes on the emulated $T_x$.

\paragraph{\B{Application Optimization:}}  Different resources may provide
different means to optimize application codes, via compiler flags, optimized
system libraries, specific hardware etc. \synapse's profiling on one system
cannot take optimization on another system into account, when those
optimizations map to different resource consumption patterns, such as GPU
acceleration which is available on the target host but was not used on the
profiling host. Profile portability is thus limited to resources with
fundamentally similar architectures.

To mitigate that affect, \synapse allows users to provide their own
application specific emulation kernels, for which they can control the
compiler flags, libraries, special instruction sets, etc. Thus, the
portability of the kernels is dependent on what was used to build/compile
them. The experiments in this paper were done with application code that was
compiled with default settings for each resource, and that uses optimized
system libraries where available.

\paragraph{\B{Multithreading:}} Application performance varies significantly
with the number of threads employed to perform the necessary operations.
While \synapse does record the number of application threads, it does not
distinguish what operations originate in what thread, nor does it use that
information during emulation (all emulation is multi-threaded though).  The
sampling based approach provides some mitigation to this, as it infers
dependencies between data and compute operations, as discussed in
Section~\ref{sec:impl}.  That inference can be wrong though, and the recorded
order of events can be a coincidence.  In that case, the sampling based
emulation will introduce too many synchronization points, and emulation will
be slower than the actual application.  This specifically can apply on
resources where resource types have very different performance
compared to the profiling machine (e.g., a much faster disks).  Whenever an
application is bound by a \I{single} resource type, that reordering effect
will not apply.

That effect can be partially mitigated by using an OpenMP emulation kernel
(the default \synapse emulation kernel for the |compute| atom supports
OpenMP), but thenumber of OpenMP threads to be used need to be configured
manually.

\paragraph{\B{Multiprocessing:}} \synapse does not attempt to detect the
spawning of additional application processes.  This could in principle be
added (|/proc/| contains the required information), but support is not
planned at this point. However, \Synapse \I{can}be used to profile and
emulate multi-process and multi-core applications: each process is handled
individually, though TCP communication is not captured (see below).

\paragraph{\B{IPC/MPI:}} \synapse does not yet profile interprocess
communication between processes within the same OS, nor any communication over
the network.  However, \I{emulation} of simple socket-based network
communication is implemented.

Specifically MPI is prevalent in the application and resource domains targeted
by \synapse.  A number of MPI tracing library exist (such as
Vampir~\cite{vampir}, the Intel MPI tracer~\cite{intel_mpi_tracer}, the IBM MPI
tracer~\cite{ibm_mpi_tracer}, dtrace~\cite{dtrace}) that could be used by
\synapse to capture the application's communication, and to replay it during the
emulation.  While we consider adding support for MPI tracing eventually (see
Section~\ref{sec:future} on future work), it will require a redesign of several
parts of \synapse, including the process management and data recording
subsystems.  An integration into \synapse is technically challenging, as
\synapse's current approach to application tracing is not able to span spawned
processes (neither local nor remote), which is a necessity to cover MPI
applications.

\paragraph{\B{Overheads:}} \synapse profiling and emulation consume certain
amounts of resources. \synapse manages though to keep those overheads very
small (see experiments in Section~\ref{sec:exp}). The profiler's start-up
time is constant and on the order of $<O(1)$ seconds. Concurrent to the
application, the profiler consumes a part of another CPU core (if available);
the profiler also uses \(\sim\)150MB of memory. Higher sampling rates can
increase the overall memory footprint of profiler. The time to write the data
to the database or to disk depends on network and disk latency and on the
total number of samples. \synapse emulation has a similar overhead (fetching
the samples from DB or disk into memory, a overhead from a tight loop that
feeds into the \synapse atoms).

The emulation additionally shows some memory overhead. This is partially owed
to the fact that multiple Python instances are spawned, and Python is
sometimes more memory heavy than the (compiled) application codes under
investigation. Though the memory overhead is not large enough to
significantly influence the \(T_x\) measured, it does show up in the profiles
of the emulation runs.

Profiling will only terminate when full sample periods have passed, which can
delay the completion of the profiling process to up to one additional
sampling period. That is only relevant for very low sampling rates.

\paragraph{\B{DB limitations:}} MongoDB has a 16MB limit on the size of a
single document. This limits the total number of data samples
\synapse supports to \(\sim\) 250,000. This limitation can be lifted by
changing to a different data model or storage backend. File-based storage of
profiles is available, which poses no limit on the number of samples.

%
\section{Experimental Results and Discussion}\label{sec:exp}

We designed our experiments to investigate the viability\mtnote{I am a bit
confused about what `viability' exactly means.}\amnote{usable, yielding the
expected results} of \synapse's approach as a
tool that: (i) automatically derives application profiles; (ii) implements
synthetic application components which can emulate the profiled applications;
and (iii) supports the tuning of such synthetic applications into various
dimensions.  The experiments 
characterize the fidelity of \synapse's profiling and emulation for a
specific scientific application, under a range of conditions 
and resources. Experiments are designed to support the requirements listed in
Sections~\ref{sec:prof_req} and~\ref{sec:emu_req} and cover the following
steps:

\begin{enumerate}
  
  \item \B{E.1 -- Profiling Overheads and Consistency:} use \synapse to profile
      an application over a range of application parameters, with different
      sampling rates.

    \B{Purpose:} determine the profiling overhead versus non-profiled execution;
    show how the consistency of profiling results depends on sampling rate.
    

  \item \B{E.2 -- Profiling Correctness and Emulation Portability:} use \synapse
      to emulate the application over the same range of application parameters,
      measuring $T_x$.

    \B{Purpose:} show the fidelity of the profiling results to capture essential
    application characteristics; support the claim that \synapse profiling
    metrics are system independent; determine emulation precision in preserving
    application $T_x$ compared to actual execution, on different resources.


      


  \item \B{E.3 -- Emulating with Different Kernels:}  use \synapse to emulate the
      same application using different kernels, measuring $T_x$, number of
      cycles used, number of instructions executed, and number of instructions
      executed per cycle.
    
    \B{Purpose:} compare the execution behavior for different emulation kernels;
    show how the choice of an application specific kernel can improve the
    fidelity of an emulation.

  \item \B{E.4 -- Emulating Parallel Execution:} use \synapse to emulate an
      application with varying degrees of single-node parallelism (OpenMP, MPI).

    \B{Purpose:} support the claim that \synapse can tune an emulated
    application in dimensions not originally supported by the actual
    application.  

  \item \B{E.5 -- Emulating Variable I/O Granularity:} use \synapse to emulate
      I/O accesses to different filesystems and with varying block sizes.
      
    \B{Purpose:} further support the claim of \synapse being able to
    fine-tune application emulation with respect to resource consumption
    details not originally profiled.

\end{enumerate}

\paragraph{Application}
The application used for the first five experiments is
Gromacs~\cite{pronk2013gromacs}. It is an application used for Molecular
Dynamics (MD) simulations, in particular for biomolecular simulations.
Gromacs is used by thousands of scientists, including multiple collaborators
of the authors.

\amnote{What physical system has been used originally?}

\mingnote{AM: Please check if the below paragraph is ok with you}
For experiments E.1--2, we configured 
Gromacs 
to run with different numbers of iteration steps, ranging from $10^4$ to
$10^7$.  The number of steps influences both CPU consumption and disk output,
but leaves disk input and memory consumption constant. Experiment E.3 is
based on the same application, but specifically focuses on the emulation of a
specific number of CPU cycles---it shows a microscopic view onto \synapse's
compute emulation capabilities. Experiment E.4 is also based on the same 
application, but focuses on the parallel emulation capabilities of \synapse.
Memory and I/O emulation is turned off for these experiments. Experiment 
E.5 uses a synthetic workloads designed to characterize \synapse's I/O 
emulation capabilities in isolation.



\paragraph{Experiment Platform} 
These experiments were performed over the course of two years, and cover a
rather wide range of resources as those were used in different projects with
\synapse-based workloads. We consider that variety as a feature, as it
demonstrates the versatility of \synapse to operate in those different
contexts. It also gives us the ability to present those experiment where
\synapse meets challenging resource variety, and where it can demonstrate how
those are managed. The downside is that the results of the experiments E.1-5
are not always directly comparable.

All profiling involving \synapse is performed on an off-the-shelf Intel Core
i7 CPU (M620) with 4 cores, 8GB memory, Intel SSD 140GB (320-Series) under a
Debian Linux with x86\_64 kernel v3.11.8-1.  That resource is named \prof in
this paper.  The E.2 emulation experiments are performed on the same host, as
well as on several HPC machines (E.3-6), namely
\stampede~\cite{stampede_details}, \archer~\cite{archer_details},
\supermic~\cite{supermic_details}, \comet~\cite{comet_details} and
\titan~\cite{titan_details}.

\stampede's compute nodes feature two 8-core Intel Xeon E5-2680 (Sandy
Bridge) processors and an Intel Xeon Phi SE10P coprocessor. We do not use the
coprocessors in our experiments. Each node has 32GB main memory and a local
250GB HDD\@. Our experiments perform all I/O on that local drive.

\archer is a Cray XC30 with two 12-core E5-2697 v2 (Ivy Bridge) series
processors and 64GB main memory per node. On \archer we also perform all disk
I/O to |/tmp|, i.e.~to a local hard drive.

\supermic's compute nodes feature two 10-core Intel Xeon E5-2680 processors
(Ivy Bridge-EP) and an Intel Xeon Phi 7120P coprocessors. We do not use the
coprocessors in our experiments. Each node has 128GB main memory. All I/O
on \supermic uses the global Lustre filesystem, unless noted
otherwise.

\comet's compute nodes feature two 12-core Intel Xeon E5-2680v3 processors.
Each node has 128GB main memory. All I/O performed on
\comet are using the NFS filesystem.

\titan is a DOE leadership class facility that currently offers the highest
degree of concurrent execution in the USA for open/academic research
(approximately 300K CPU cores).  Its compute nodes feature a 16-core AMD
Opteron 6274 CPU with 32 GB of DDR3 ECC memory and an Nvidia Tesla K20X GPU
with 6 GB GDDR5 ECC memory.  Unless noted otherwise, all experiments use the
Lustre filesystem for all I/O.

The data sets produced and used in the experiments as presented below are
freely available, as is the \synapse software itself; see the \I{``Software
Availability''} paragraph in Sec.~\ref{par:software}.

\subsubsection*{Experiment E.1: Profiling Overheads and Consistency}

\begin{figure}[!t]
 \centering
  \includegraphics[width=0.85\linewidth]{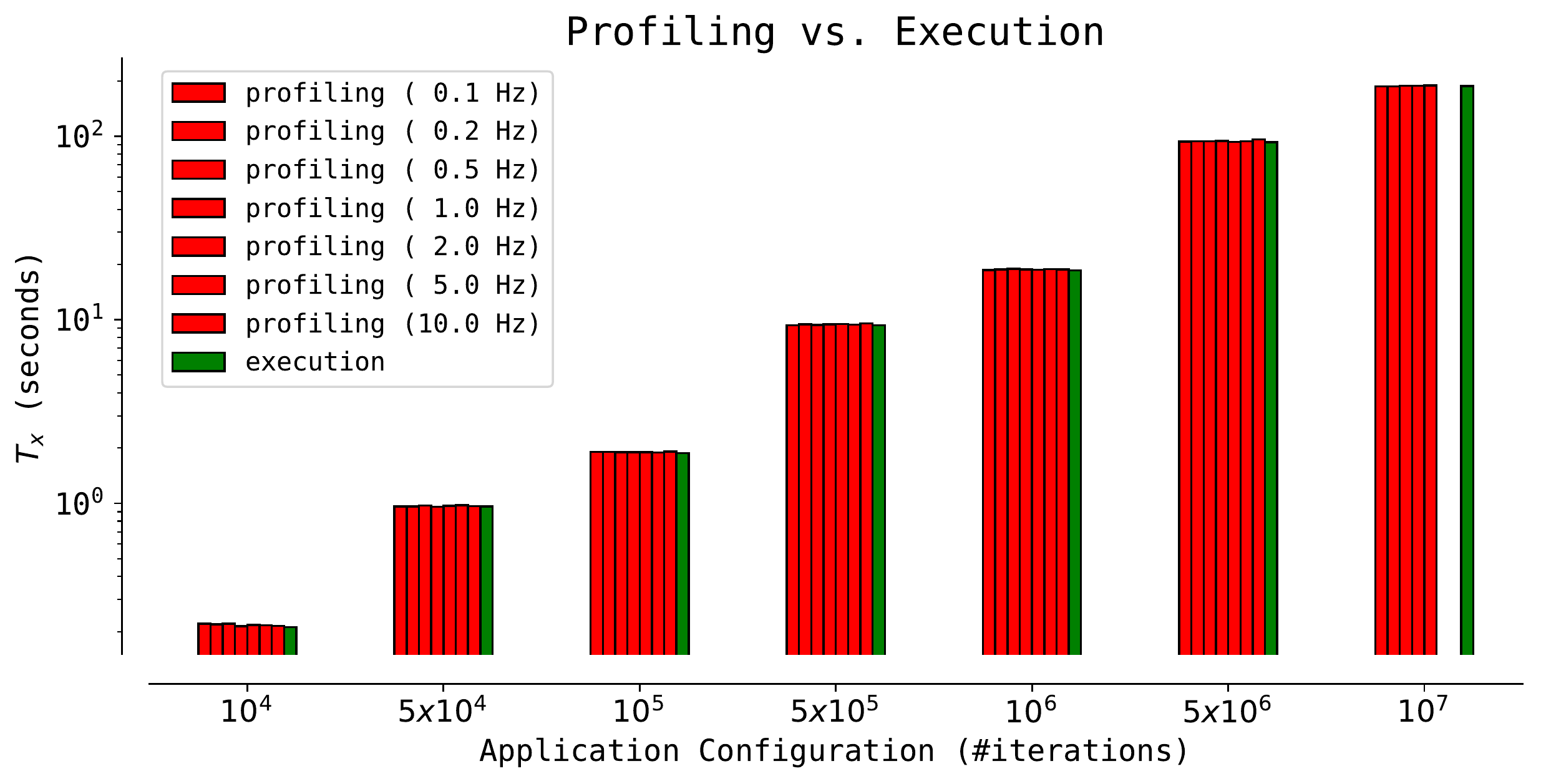}
  \caption{
      \footnotesize
      \B{Profiling Overhead:} Profiling consumes additional system
      resources, but does not affect the $T_x$ of the application profiled.
      \label{fig:pro_overhead}
  }
\end{figure}

\begin{figure}[!b]
 \centering
  \includegraphics[width=0.85\linewidth]{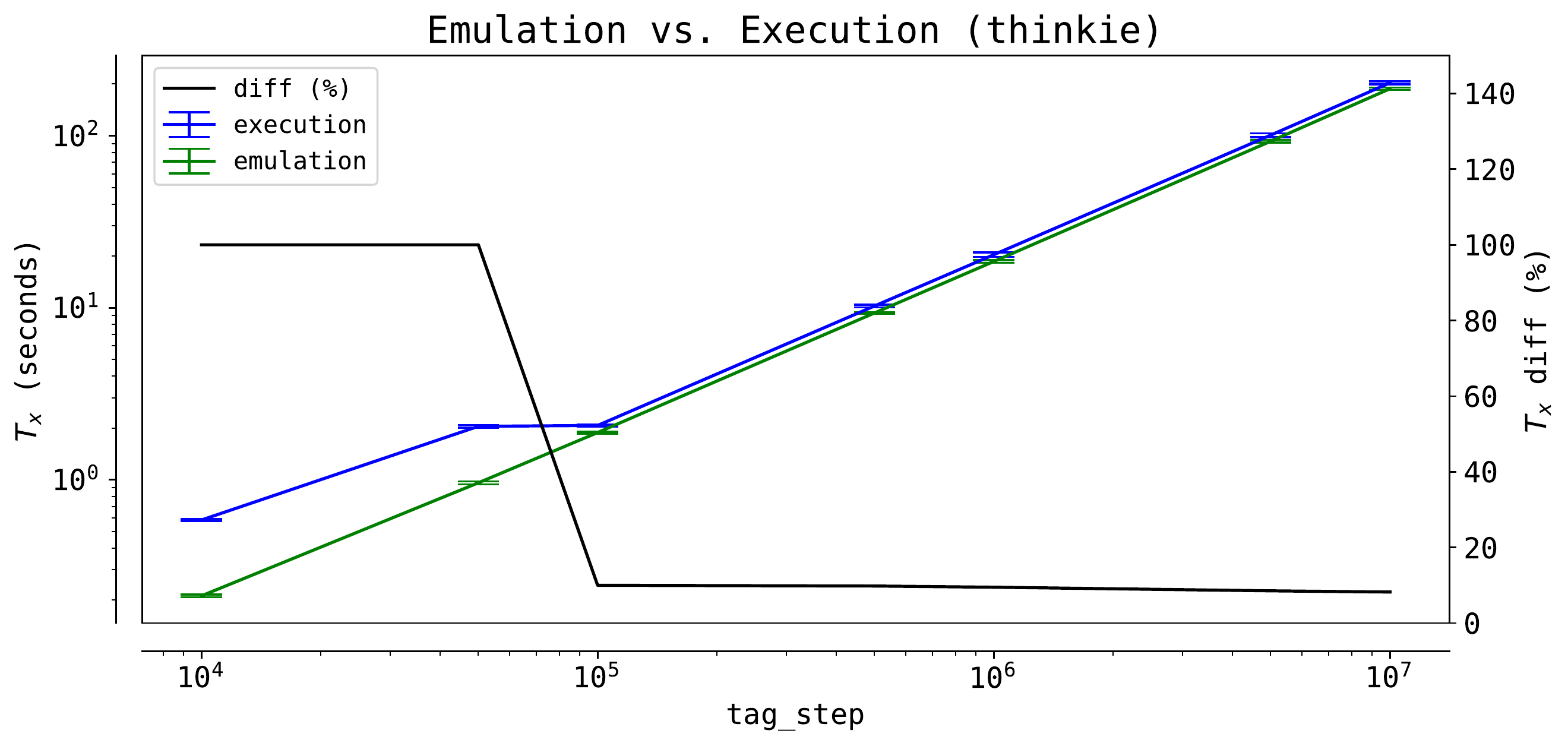}
  \caption{
      \footnotesize
      \B{Emulation Correctness:} when using the same resource,
      emulated runtimes (green) agree with actual application runtimes (blue)
      for runtimes larger than the \synapse startup delay ($\sim 1 sec$).
      \label{fig:emu_correctness}
  }
\end{figure}

\begin{figure}[!t]
 \centering
 \includegraphics[width=0.85\linewidth]{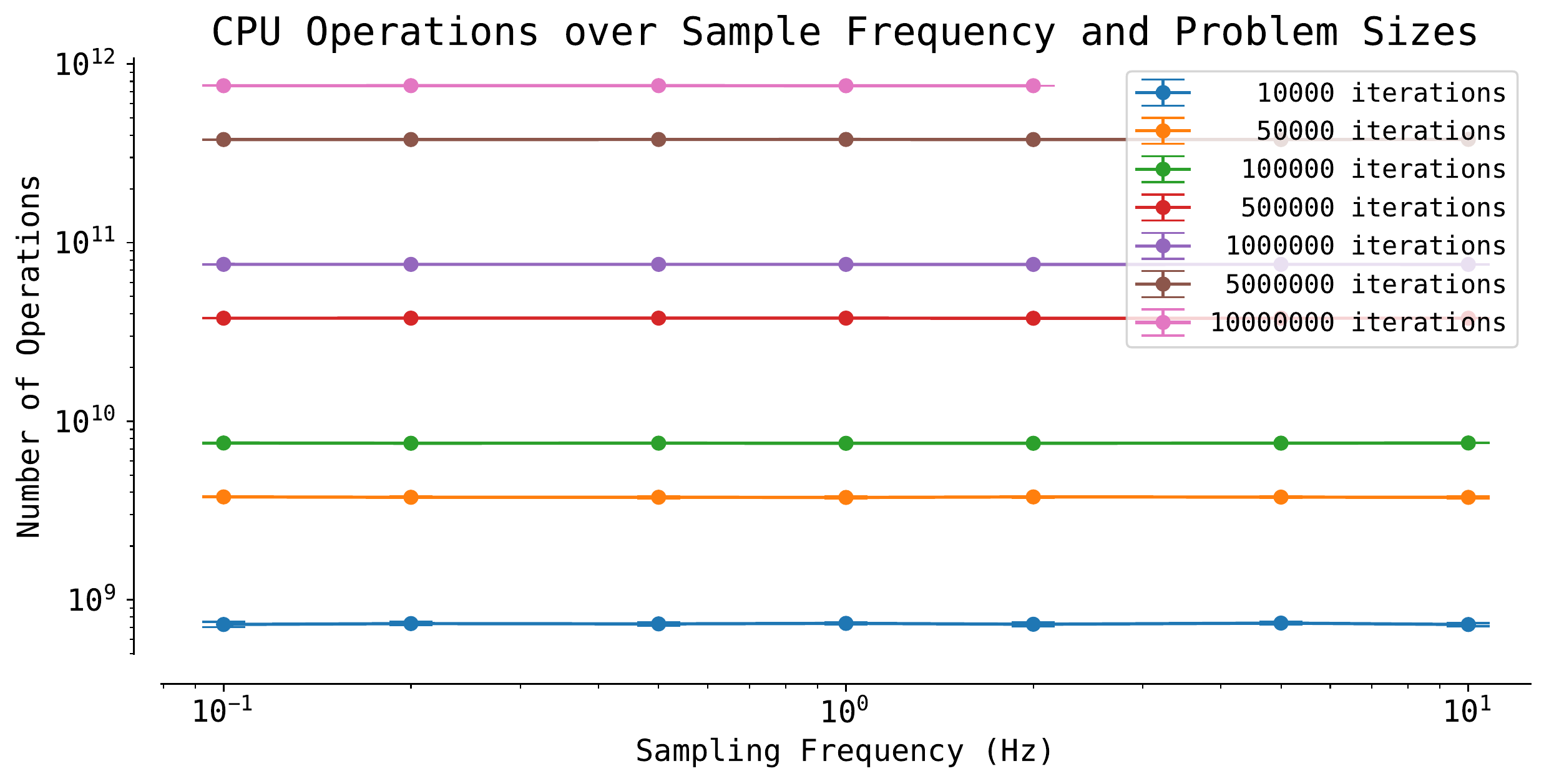}
 \includegraphics[width=0.85\linewidth]{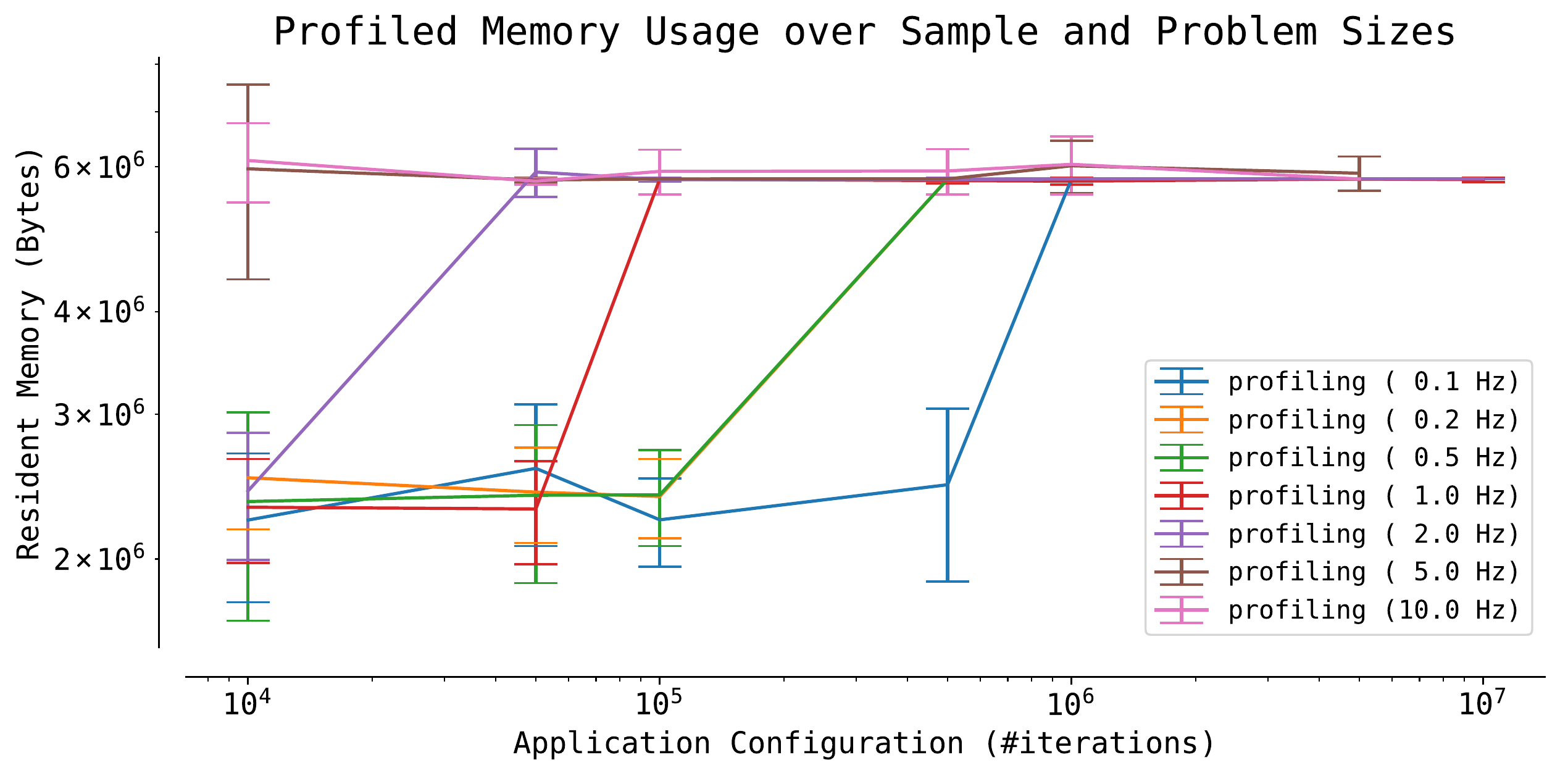}
  \up
  \caption{
      \footnotesize
      \B{Profiling Consistency (top):} Independent of the
      profiler sampling rate, \synapse reports very consistent values for
      consumed CPU operations, for a wide range of application $T_x$
      (log/log scale, the plot includes error bars).
      \B{(bottom):} For some metrics, the profiler
      requires sample rates to be smaller than application runtime. For the
      example here (resident memory), the measure is underestimated by the
      profiler for sample rates that allow only one data sample to be taken
      over the course of the application runtime. For multiple samples, the
      measures quickly stabilize.
      \label{fig:pro_consistency}
  }
\end{figure}

\paragraph{Profiling Self-Inference and Overhead}
Figure~\ref{fig:pro_overhead} compares $T_x$ for two cases: native
application runs, and application execution while using the the \synapse
profiler with different sampling rates.

The graph shows negligible profiling overhead for the
investigated range of problem sizes and sampling rates: 
the runtime is constant for all application configurations, independent of the sampling
rate. The largest configuration misses one data sample due to limitations in
the database backend (see~\ref{sec:limits}).

\paragraph{Profiling Consistency}
We repeated profiling of the same application instances in the same
environment.  While the non-zero standard deviation indicates some noise in
the measured metrics, the distribution is in very good agreement with the
distribution of the pure application $T_x$ (see
Figure~\ref{fig:pro_consistency}), which indicates the influence of system
background.  The figure shows the profiling consistency over a range of
application sizes and sampling rates.

Please note that this experiment to a large part confirms the consistency of
the underlying profiling tools used by \synapse---apart from correct data
collection and interpretation, \synapse by design does not influence that
level of consistency.

\subsubsection*{Experiment E.2: Profiling Correctness and Emulation Portability}

The ultimate purpose of \synapse's profiling is to feed \synapse's emulation, to
emulate the profiled application on any target resource.
Figure~\ref{fig:emu_correctness} compares the $T_x$ of the application
execution to $T_x$ of emulated application runs, on the same resource as used
for profiling.  The graph shows that emulation tends to incur an overhead,
specifically at startup time, which quickly becomes insignificant for
applications running longer than the \synapse Emulator startup delay ($\sim
1 sec$),

As a sanity check, we profiled the \I{emulated} application and 
compared the reported system resource consumption results: the 
values are in excellent agreement for any application running
longer than a few seconds, when the sample rate is fast enough
to result in at least two samples. There are some small 
deviations due to the memory footprint of the emulation driver 
(Python, C threads), but no other discernible difference otherwise 
(graph is not included).
\amnote{if the above paragraph is considered too vague, we should remove it.}

\begin{figure}[!t]
    \centering
    \includegraphics[width=0.85\linewidth]{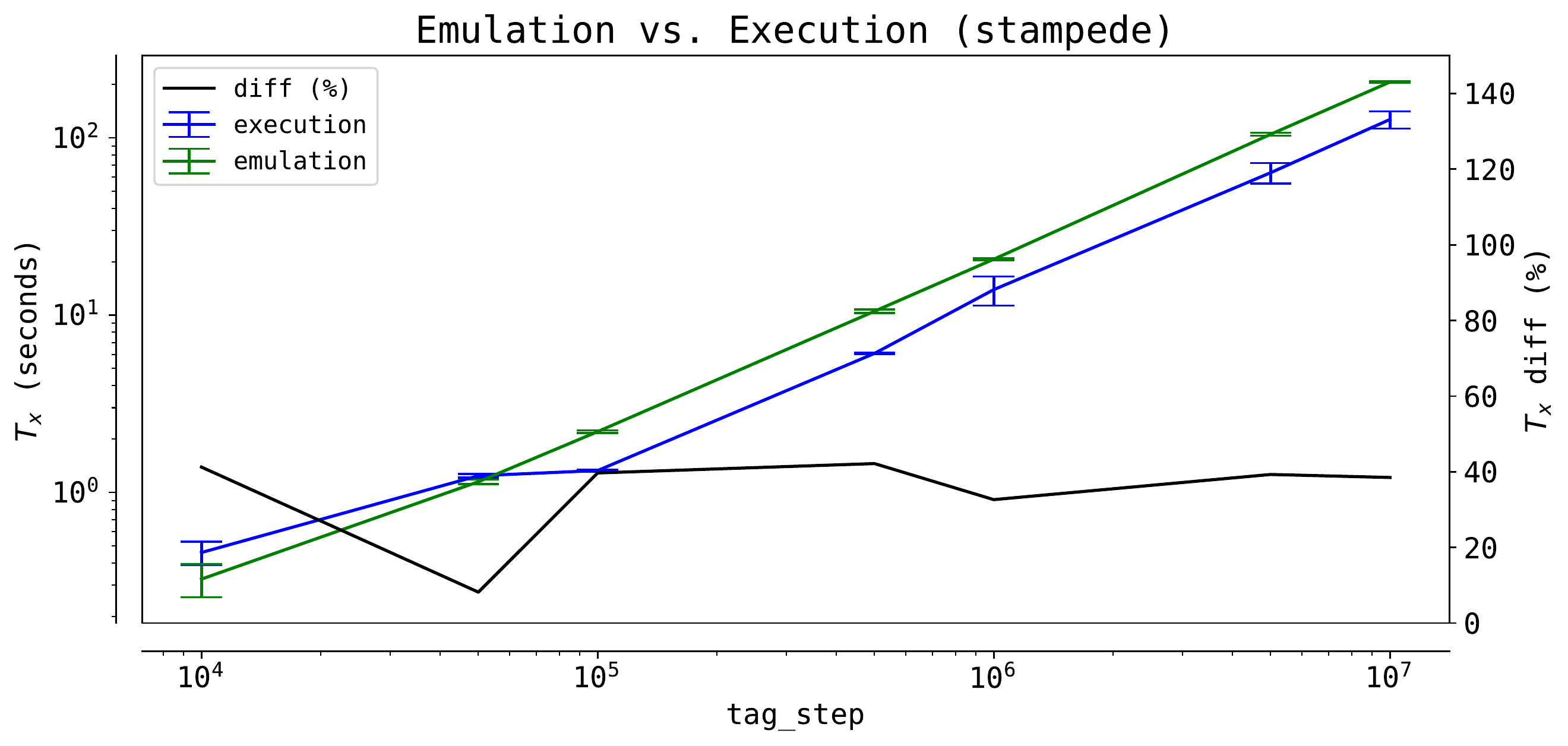}
    \includegraphics[width=0.85\linewidth]{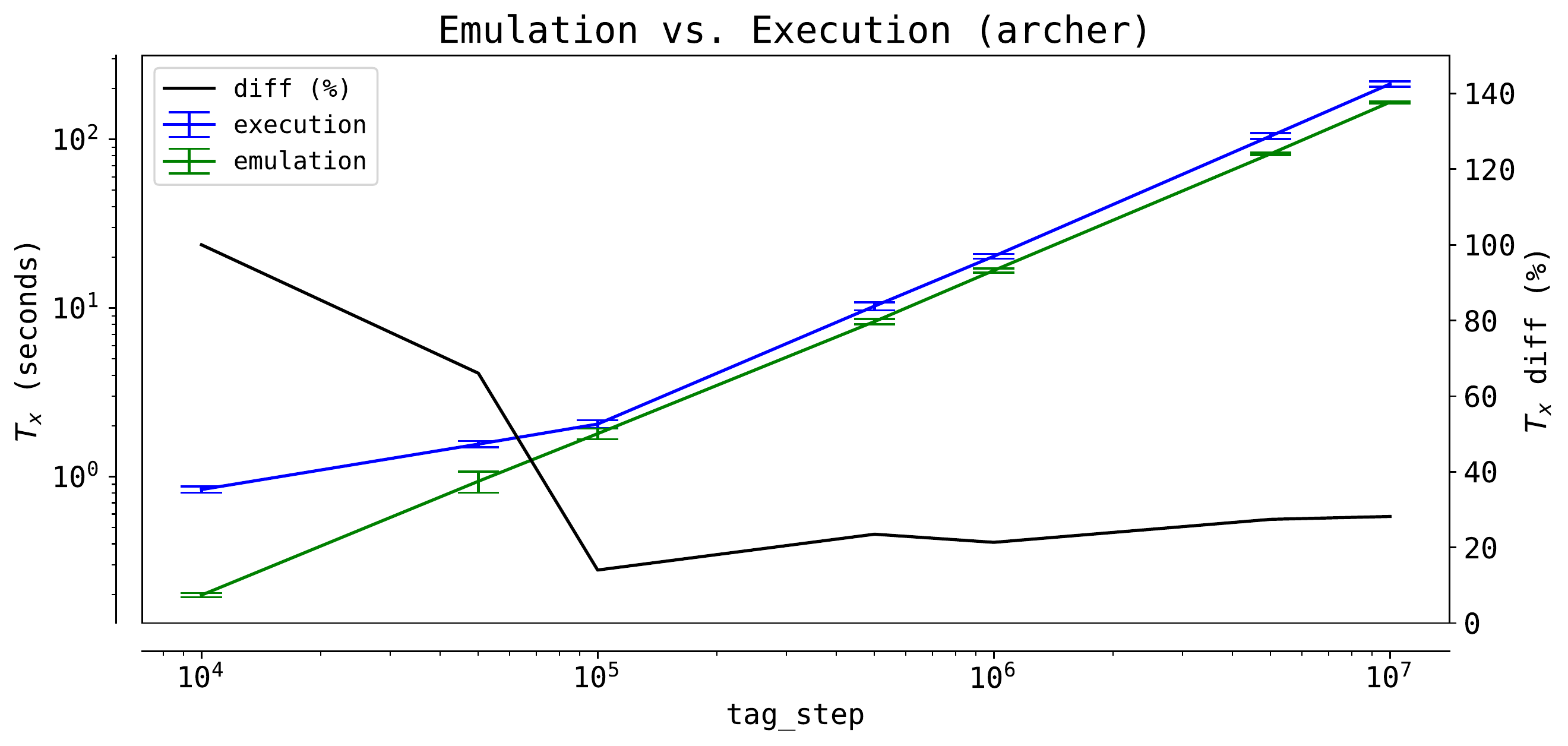}
    \caption{
        \footnotesize
        \B{Emulation Correctness:} comparison of $T_x$ for the originnal
        application and its emulation running on resources \I{different} than
        the one used for profiling: \stampede (top) and \archer (bottom). The
        second y-axis (right side) shows the difference between application and
        emulation in $\%$. }
    \label{fig:emu_cor}
\end{figure}

Figure~\ref{fig:emu_cor} compares application execution and application
emulation on resources \I{different} than the one used for profiling,
specifically on \stampede and \archer{}. Over the investigated range of
application sizes and sampling rates, the $T_x$ of the application and its
emulation resemble the essential application's execution characteristics
measured on the original resource. Again, the plots show that the emulation
overhead is significant for very 
short runtime.

The emulation on \stampede in Figure~\ref{fig:emu_cor}~(top) is consistently
\I{faster} compared to the application execution. The difference between
application emulation converges to $\sim 40\%$.  The emulation $T_x$ on
\archer in Figure~\ref{fig:emu_cor}~(bottom) is consistently \I{slower} than
the $T_x$ of the actual application---the difference converges to $\sim
33\%$.

While the overall trend of the $T_x$ development is well captured, the
absolute values for $T_x$ differ significantly for both machines.  For the
use cases discussed, the preciseness of the emulation is much less of a
concern than the ability to reproduce scaling behavior.  Nevertheless, the
next experiment will discuss and address that discrepancy for those use cases
where preciseness is an important concern.

\subsubsection*{Experiment E.3: Emulating with Different Kernels}

\begin{figure}[!t]
    \centering
    \includegraphics[width=0.85\linewidth]{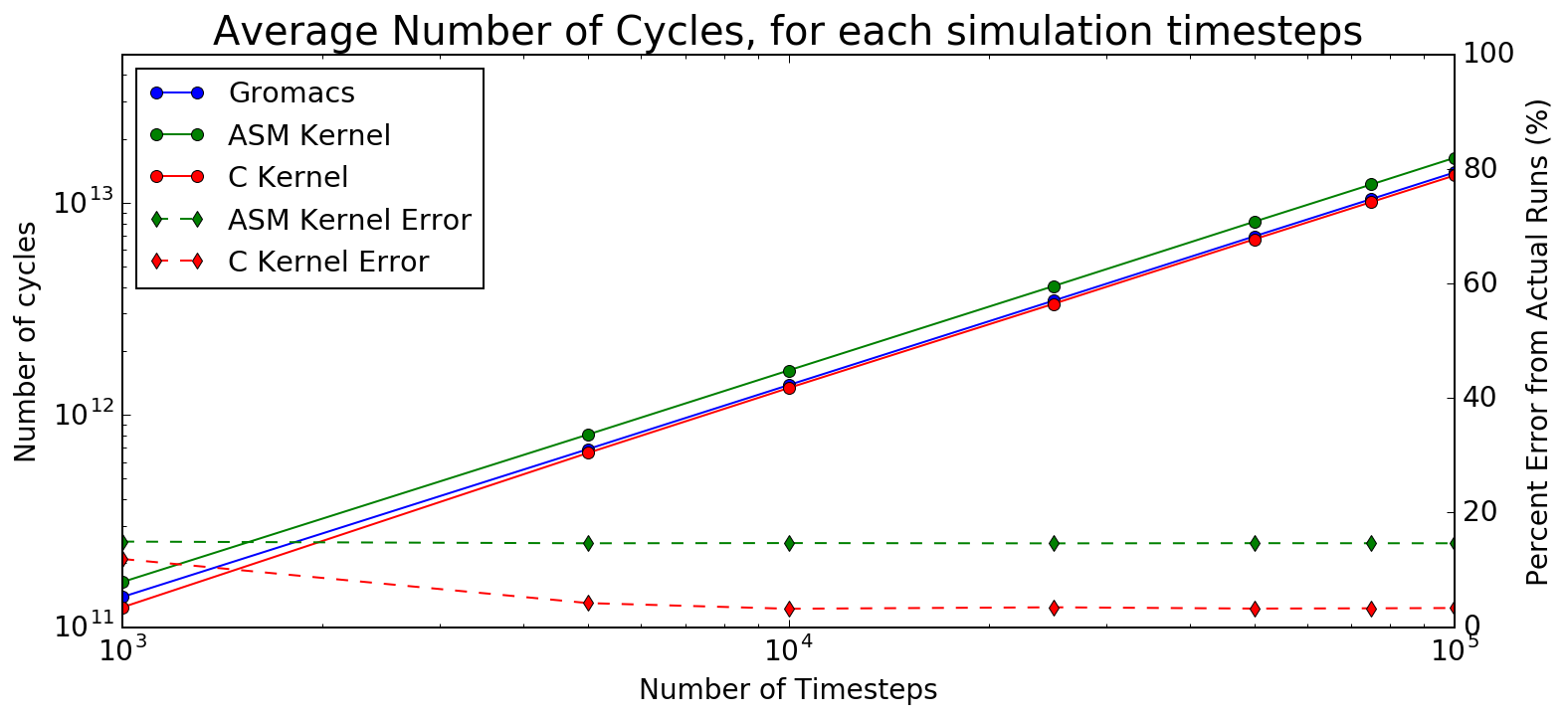}
    \includegraphics[width=0.85\linewidth]{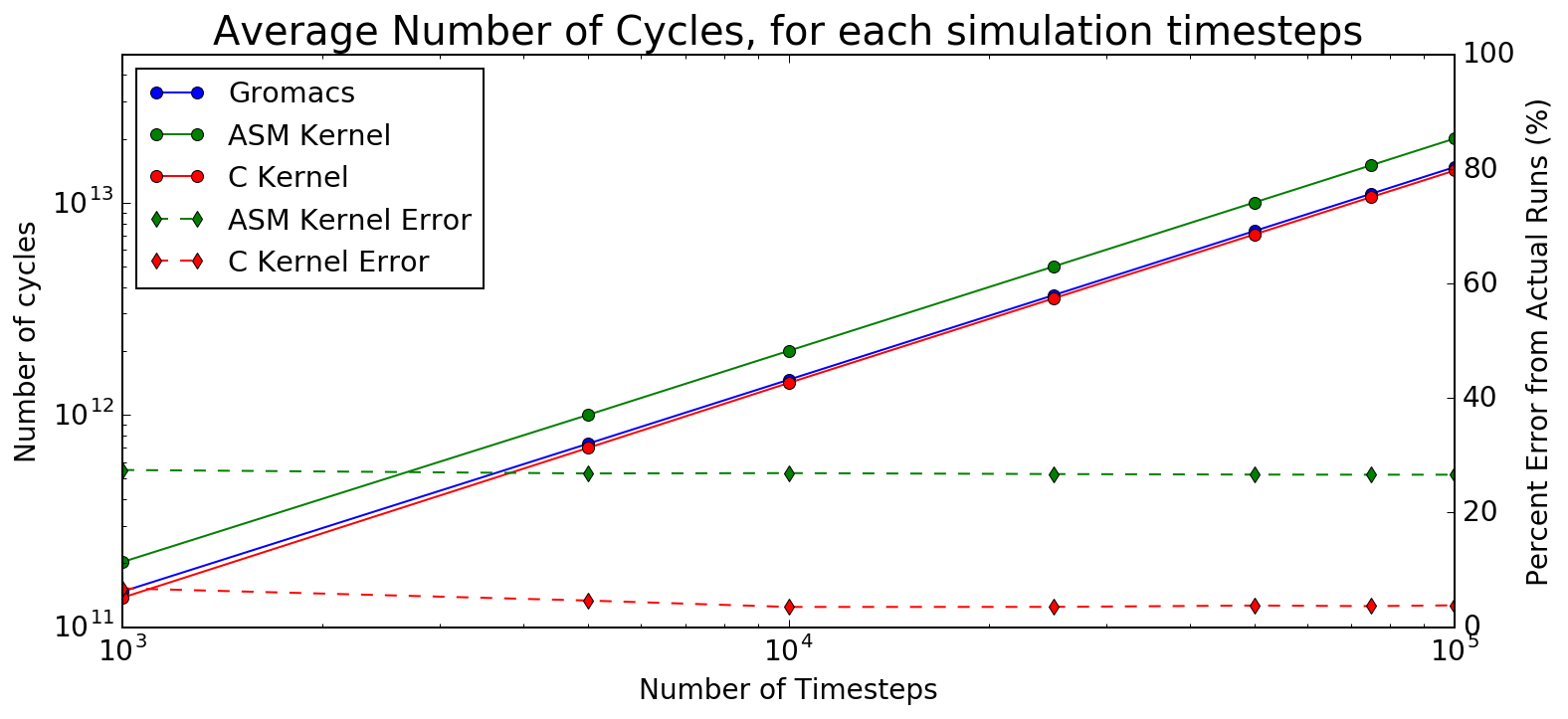}
    \caption{\footnotesize
        The number of cycles used by the Gromacs application and emulation
        with the C and ASM matrix multiplication kernels.  The second Y-axis
        shows the error percentage of the emulation when compared to the
        application run (the absolute error bars are too small to be
        visible).  Data have been measured on \comet (top) and
        \supermic (bottom).
    }\label{fig:exe_emu_diff_cy}
\end{figure}

\begin{figure}[!t]
    \centering
    \includegraphics[width=0.85\linewidth]{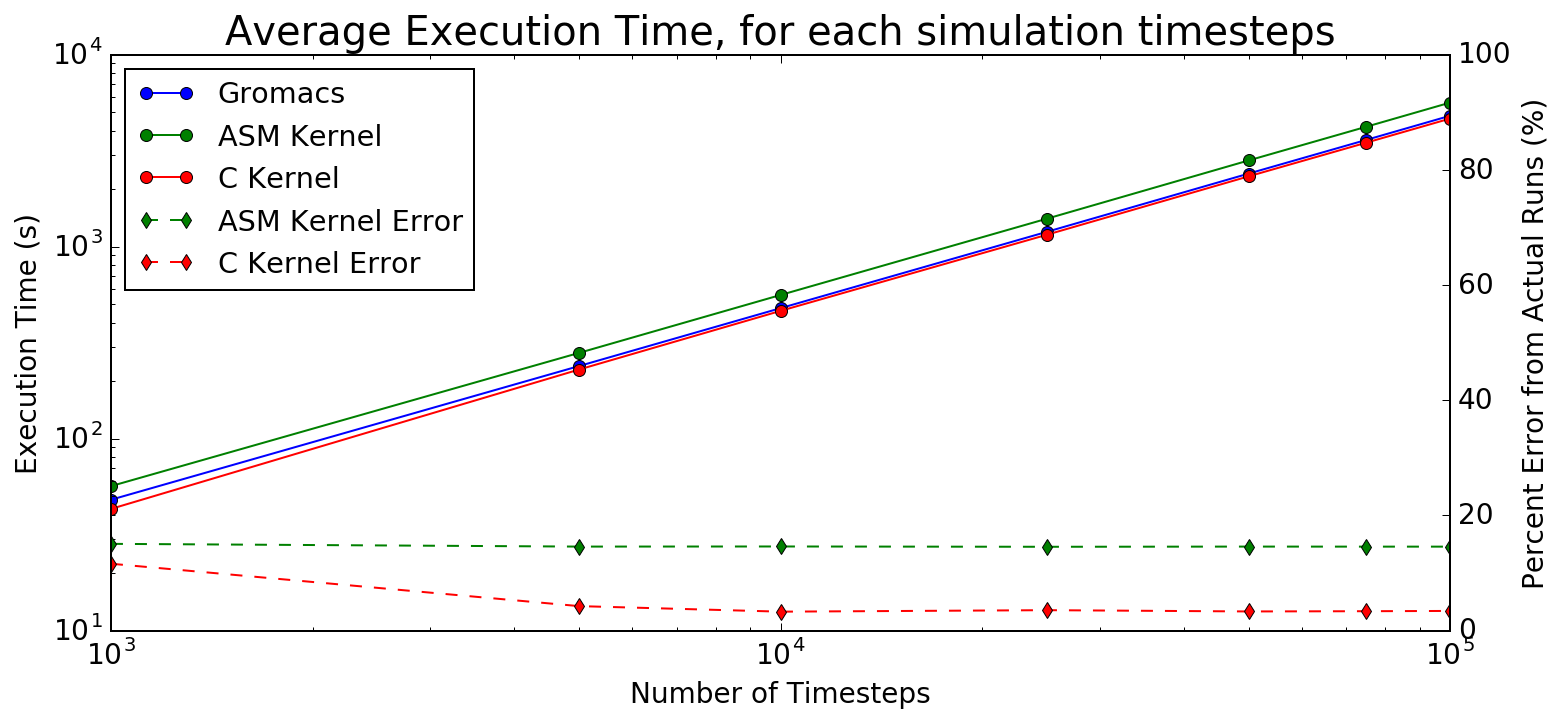}
    \includegraphics[width=0.85\linewidth]{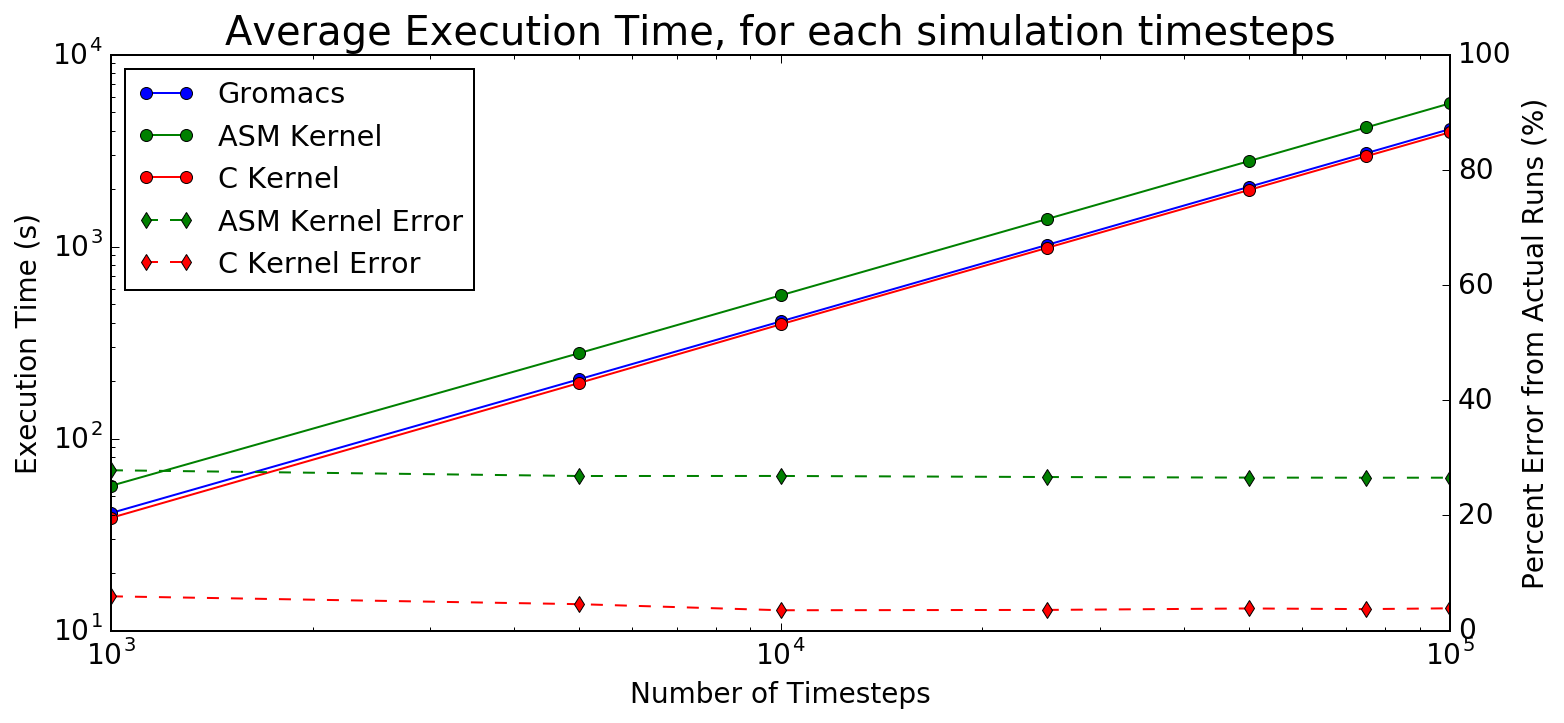}
    \caption{
      \footnotesize
      The $T_x$ of the Gromacs application and emulations using
      the C and ASM kernels, on \comet (top) and \supermic (bottom). 
      The second Y-axis again shows the emulations' error percentage.
    }\label{fig:exe_emu_diff_tx}
\end{figure}

\begin{figure}[!t]
    \centering
    \includegraphics[width=0.85\linewidth]{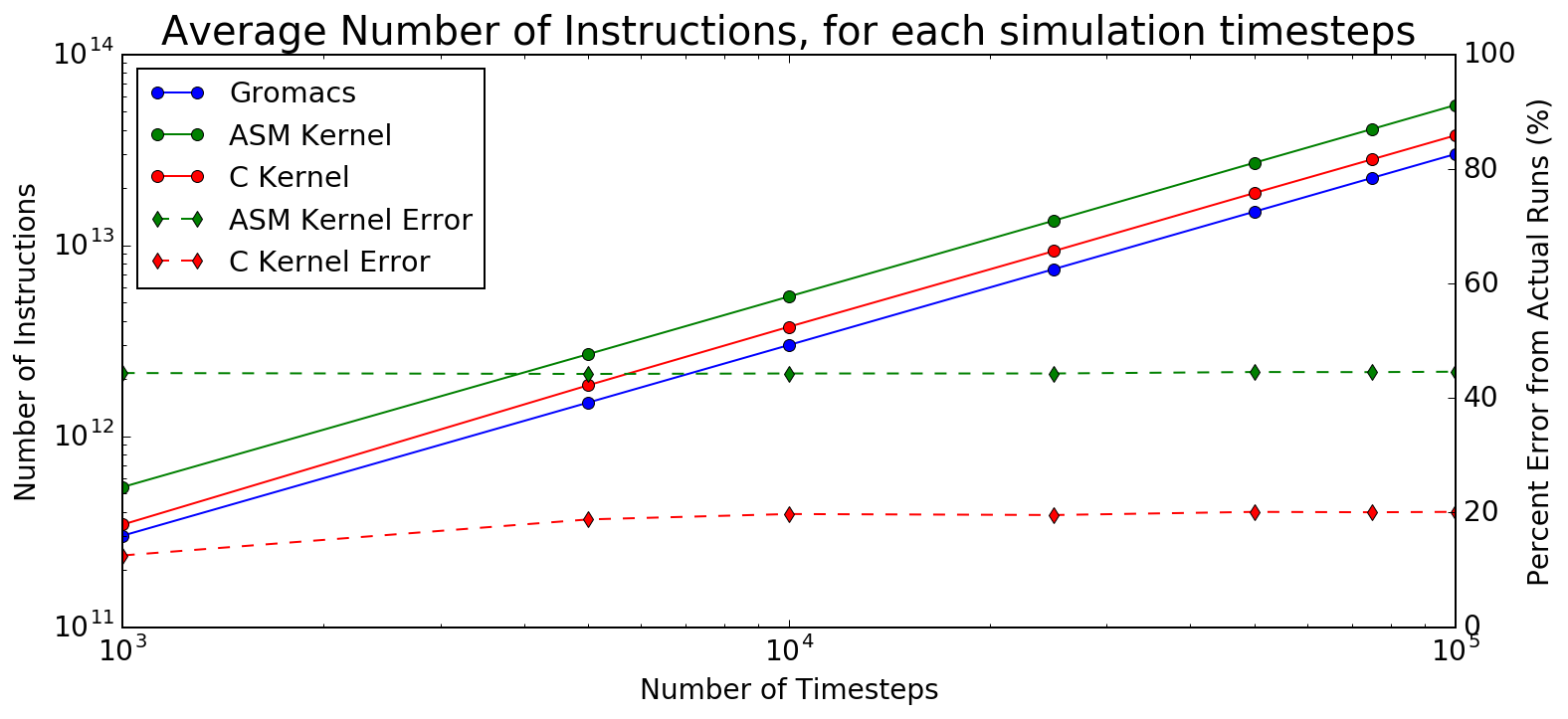}
    \includegraphics[width=0.85\linewidth]{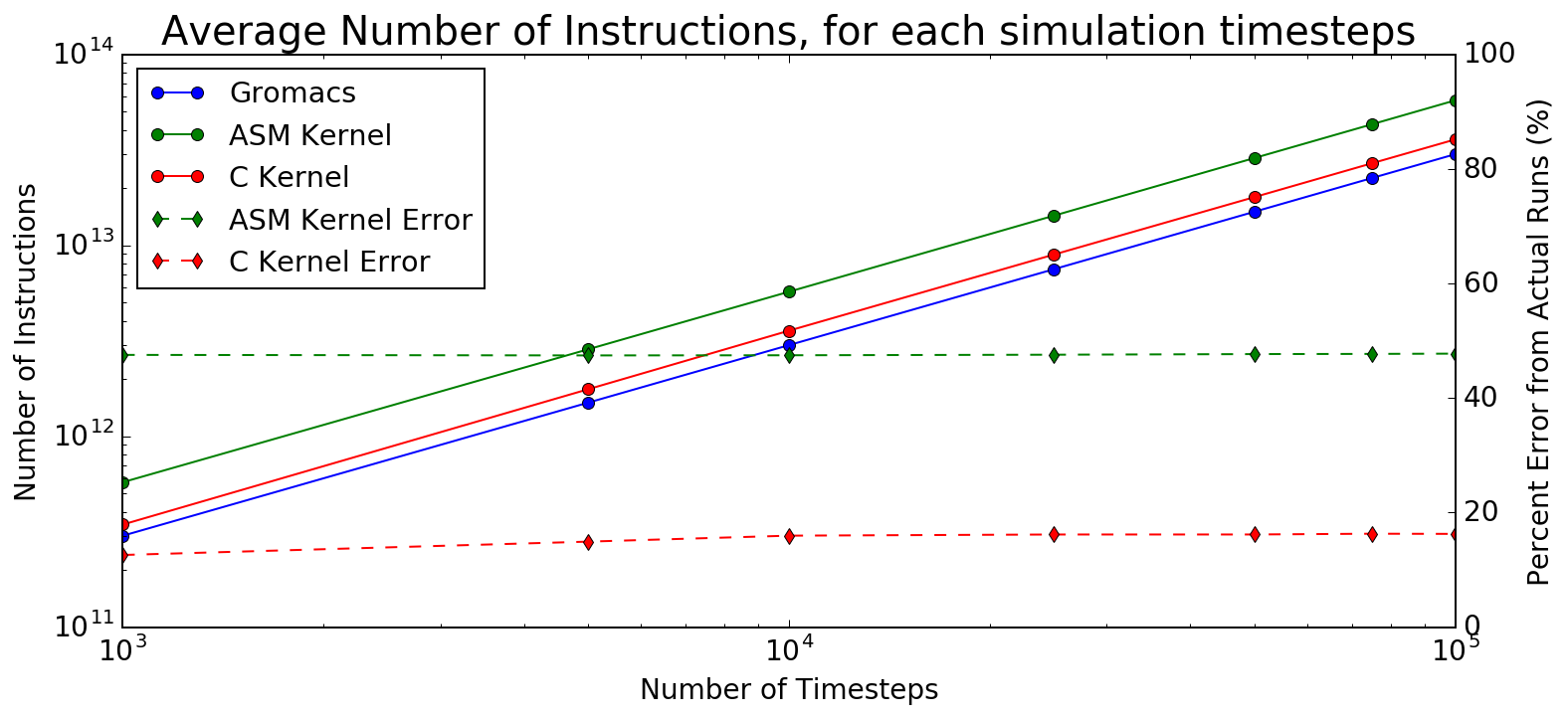}
    \caption{
      \footnotesize
      The number of instructions executed by the Gromacs application and the
      emulations, using both the C kernel and ASM kernel on \comet (top) and
      \supermic (bottom).  The second Y-axis shows the emulations' error percentage.
    }\label{fig:exe_emu_diff_instr}
\end{figure}

\begin{figure}[!t]
    \centering
    \includegraphics[width=0.85\linewidth]{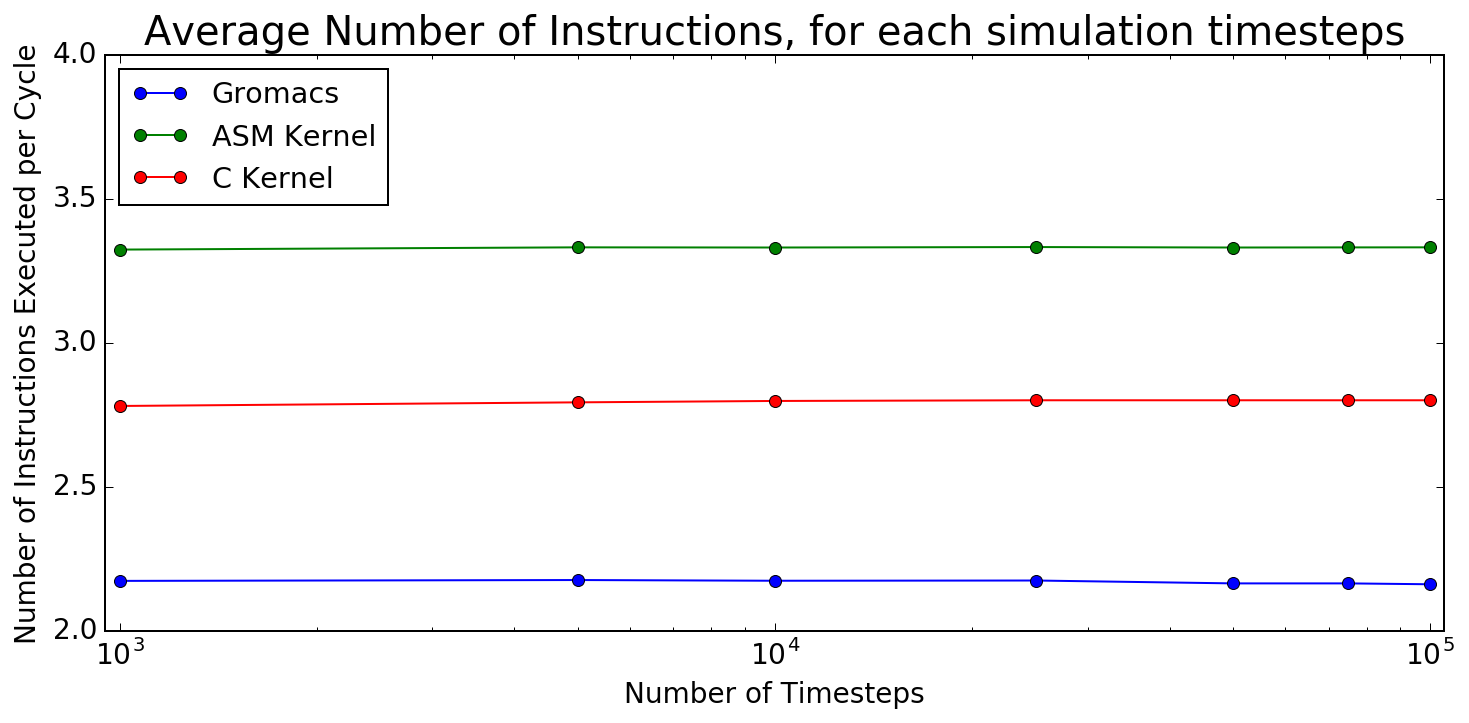}
    \includegraphics[width=0.85\linewidth]{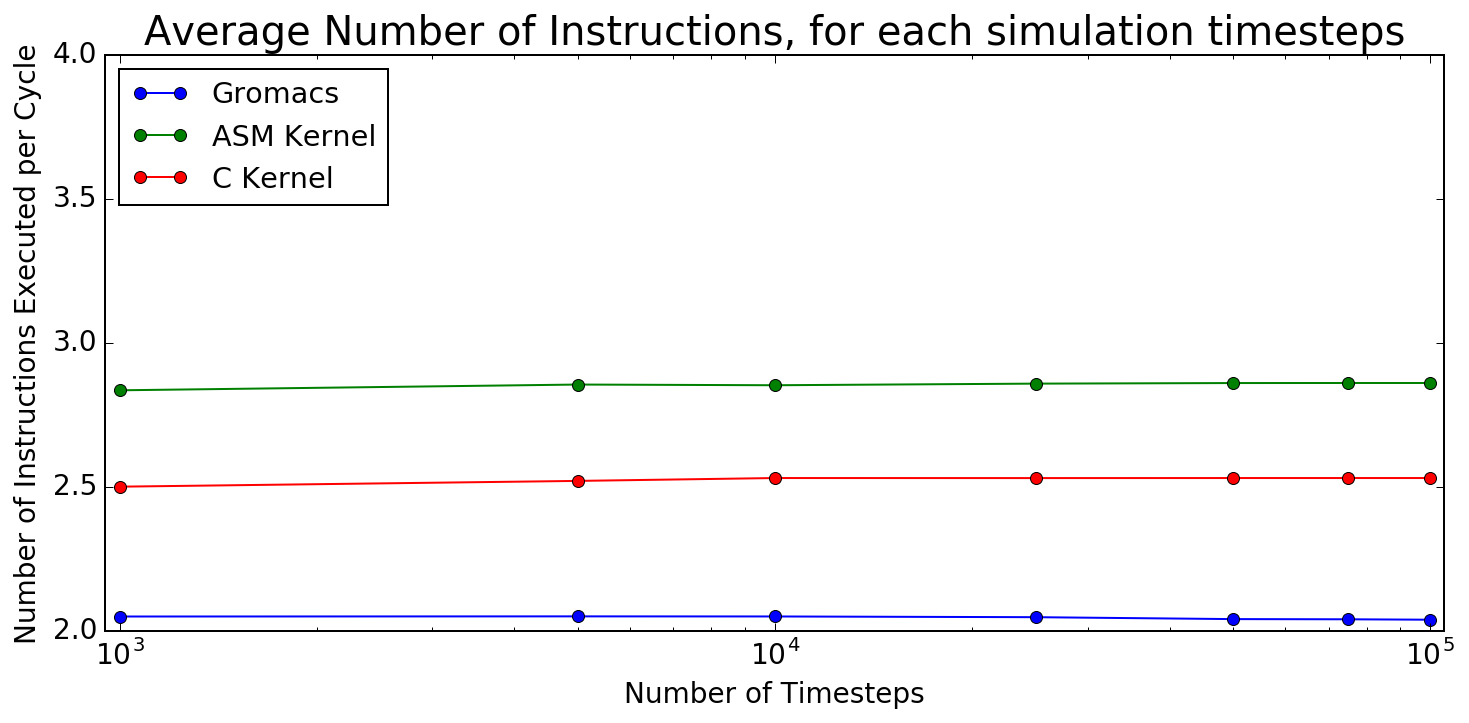} 
    \caption{
      \footnotesize
      Number of instructions executed per cycle (`instruction
      rate') for Gromacs application and emulations, measured again on \comet
      (top) and \supermic (bottom).
    }\label{fig:exe_emu_diff_instr_rate}
\end{figure}

We investigated how using different kernels affects the fidelity with which
\synapse emulates the execution of an application. Using again Gromacs as the
application, we emulated the execution of Gromacs by directing \synapse to
consume the same number of cycles as consumed by Gromacs when running a given
number of iteration steps. We emulated the cycle consumption using either a
naive matrix multiplication kernel implemented in C or a matrix
multiplication kernel implemented using x86 assembly (ASM). The primary
difference between the two kernels is that the matrices used by the (ASM)
kernels fit in cache, while the matrices used by the C kernel does not. Thus,
the two kernels have different memory access patterns when performing matrix
multiplication.  While we did not specifically investigate the caching
behavior of Gromacs, we would naively expect the C kernel represent Gromacs'
memory access patterns more truthfully.

We first profiled a set of sample runs to investigate the execution behavior
of Gromacs and measured the average number of cycles required to execute a
simulation. We also measured $T_x$, the total number of instructions
executed, and average instruction rate. We profiled Gromacs simulations
running 1,000, 5,000, 10,000, 25,000, 50,000, 75,000 or 100,000 iterations.
Then, we emulate the simulation using both the C kernel and the ASM kernel to
consume some number of cycles equal to the average number of cycles used by
the Gromacs application to run a given number of iterations. We also profiled
those emulation runs to collect the same information as we did with the
original application. This allows us to compare the fidelity of the emulation
relative to the execution of the original application. The application and
their associated emulations were executed on \comet and \supermic.

Figures~\ref{fig:exe_emu_diff_cy},~\ref{fig:exe_emu_diff_tx},~\ref{fig:exe_emu_diff_instr}
and~\ref{fig:exe_emu_diff_instr_rate} respectively show the average number of
cycles used, $T_x$, number of executed instructions, and instruction rate, of
the Gromacs application and its emulations on \comet and \supermic.
Figures~\ref{fig:exe_emu_diff_cy},~\ref{fig:exe_emu_diff_tx}
and~\ref{fig:exe_emu_diff_instr} also show the respective error percentage when
comparing emulation via the C and ASM matrix multiplication kernels to the
application runs.  All data points are plotted with error bars denoting a 99\%
confidence interval; for all data points, the width of the confidence interval
is no more than 6.6\% of the value of the data point.

Figure~\ref{fig:exe_emu_diff_cy} investigate the relation between the average
number of cycles used by Gromacs simulations and by the associated emulations
for each number of iterations given above.  The error percentage on \comet
(top) converges to $\sim 3.5\%$ for the C kernel, and to $\sim 14.5\%$ for
the ASM kernel.  For \supermic (bottom), the C kernel error converges to $\sim
4.0\%$, the ASM kernel error converges to $\sim 26.5\%$.

For all data points, the error percentage of the C kernel emulation is
smaller than that of that of the ASM kernel emulation. Moreover, we find that
the confidence interval of the average number of cycles used by emulations is
three order of magnitude smaller than the corresponding average. This implies
that the number of cycles used when directed to consume some fixed number of
cycles is consistent between runs.

Figure~\ref{fig:exe_emu_diff_tx} shows that applying the C kernel with its
more precise emulation of used CPU cycles results in a much more precise
emulation of $T_x$ over the full investigated range of parameters when
compared to the ASM kernel: on \comet, tHe error percentage of the C kernel
converges to $\sim 3.5\%$, while for the ASM kernel it converges to $\sim
14.5\%$.  On \supermic, the error percentages converge to $\sim 4.0\%$ and
$\sim 26.5\%$ for the C and ASM kernels, respectively.

We note also that the error percentage for emulations using a given kernel is
very similar to the error percentage of the number of cycles for emulations
using the same kernel. This is consistent with Gromacs simulations being
`compute-heavy', i.e.~, the computational load being the dominant factor for
$T_x$, and with the fact that the average clock speeds measured when
executing the original application and the emulations were very consistent
($\sim$2.88-2.90 GHz on \comet, $\sim$3.58-3.60GHz on \supermic). We thus can
approximate the $T_x$ of a Gromacs simulation as the quotient of the number
of cycles required to perform the computation and the clock speed measured.

Figures~\ref{fig:exe_emu_diff_instr} and~\ref{fig:exe_emu_diff_instr_rate}
provide additional insight into the behavior of the Gromacs simulation and
emulations.  The error percentages for the executed number of instructions
converge to $\sim 20.0\%$ (\comet) and  $\sim 16.0\%$ (\supermic) for the C
kernel, while for the ASM kernel they converge to $\sim 44.5\%$ (\comet) and
$\sim 47.5\%$ (\supermic), respectively. Not only does the C kernel better
reproduce the number of consumed CPU cycles, it also better emulates the
number of instructions executed per cycle:
Figure~\ref{fig:exe_emu_diff_instr_rate} shows for \comet (top) a measured
rate of instructions of $\sim 2.17/cycle$ for the Gromacs application run,
and rates of $\sim 2.80/cycle$ and $\sim 3.30/cycle$ for the C and ASM
kernels, respectively; for \supermic (bottom), we measure a rate of $\sim
2.04/cycle$ for the application execution, while finding rates of $\sim
2.53/cycle$ and $\sim 2.86/cycle$ for the C and ASM kernels, respectively.

The data thus show that the number of cycles
(Figure~\ref{fig:exe_emu_diff_cy}), execution time
(Figure~\ref{fig:exe_emu_diff_tx}) the number of instructions
(Figure~\ref{fig:exe_emu_diff_instr}) and the rate of executing those
instructions (Figure~\ref{fig:exe_emu_diff_instr_rate}) are better reproduced
by the C kernel than by the ASM kernel. That the measured instruction rates
for the C kernel is lower than for the ASM kernel matches our expectation as
we do not expect the matrices used in matrix multiplication operations to
always fit into cache.

To summarize: of the two kernels used to emulate the Gromacs simulation, the
execution behavior of emulations using the C kernel is more similar to that
of the actual application, for all investigated parameters and resources.
This supports our claim that users can control the manner in which \synapse
consumes system resources (e.g., cycles), by using application specific
emulation kernels, thus increasing the fidelity with which applications are
emulated, where that high fidelity is required.

Implementing application specific kernels requires knowledge of the
application or understanding of the profiler data measured for that
application. To reduce the need of the former, we plan to provide better
support for the latter by extending \synapse so that it is able to interface
with other, more sophisticated profilers: a better characterization of
similarities between emulation kernels and applications will inform users on
how to develop kernels that emulate applications with greater fidelity. We
provide specialized kernels for applications related to our own research
(incl. Gromacs and Amber). The default ASM kernel though will continue to
form a viable baseline, offering a reasonable application representation
where fidelity requirements are not as stringent, while providing good
tuneablity and scalability.



\subsubsection*{Experiment E.4: Emulating Parallel Execution}

\Synapse's profiling capabilities for parallel applications are, as
discussed, rather limited.  However, several of our motivating use cases call
for the controlled execution of parallel tasks.  \Synapse does offer the
ability to distribute a workload via either OpenMP (multi-threading) or
OpenMPI (multi-processing).  That parallelism does only affect the compute
emulation though - all other metrics are handled in the naive way (resource
sharing in the case of multi-threading, duplicated resource usage in the case
of multi-processing).  Also, \synapse at this point makes no attempt to
emulate any communication, which for many applications can be a dominating
factor.

However, even with those fundamental limitations and constraints, we include
this experiment to document preliminary results, showing that the parallel
execution can already be used to emulate more heterogeneous workloads than
single-core applications.  This turns out to be particularly useful and
relevant when used in the context of HPC and HTC middleware development,
where single-core tasks are more the exception than the rule.

\begin{figure}[!t]
  \centering
  \includegraphics[width=0.85\textwidth]{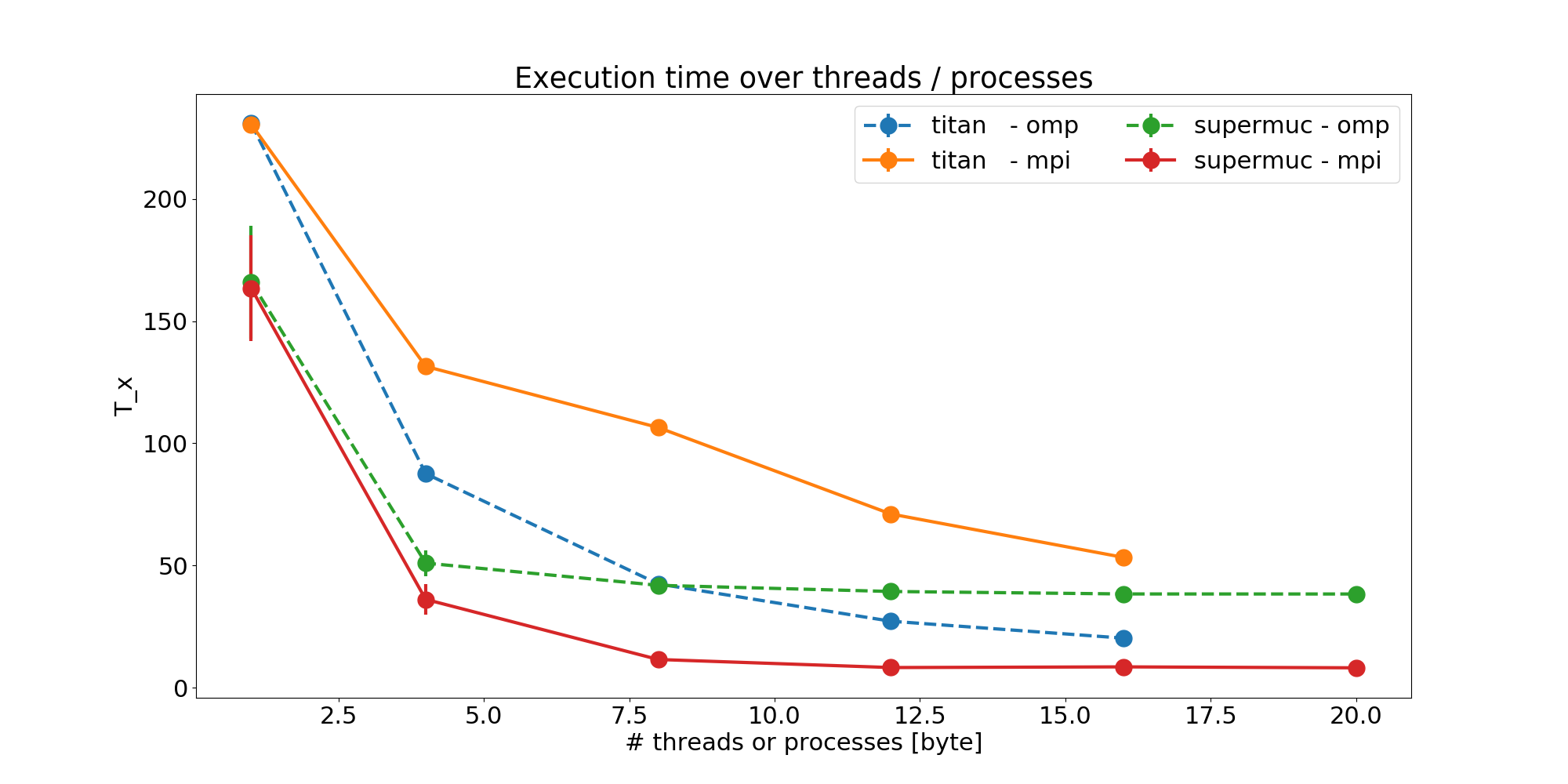}
  \caption{\textbf{Application Concurrency:} this experiment applies thread
    (OpenMP) or process based (OpenMPI) parallelism to a profile obtained
    from a single-threaded application run.  The resulting scaling plots on
    \titan and \supermic show a behavior similar to what the actual parallel
    application execution also yields: good scaling for small core numbers,
    but diminishing return for larger core numbers, where overall system
    stress limits potential performance gains.\label{fig:omp}
  }
\end{figure}

\begin{figure}[t!]
    \centering
    \includegraphics[width=0.9\linewidth]{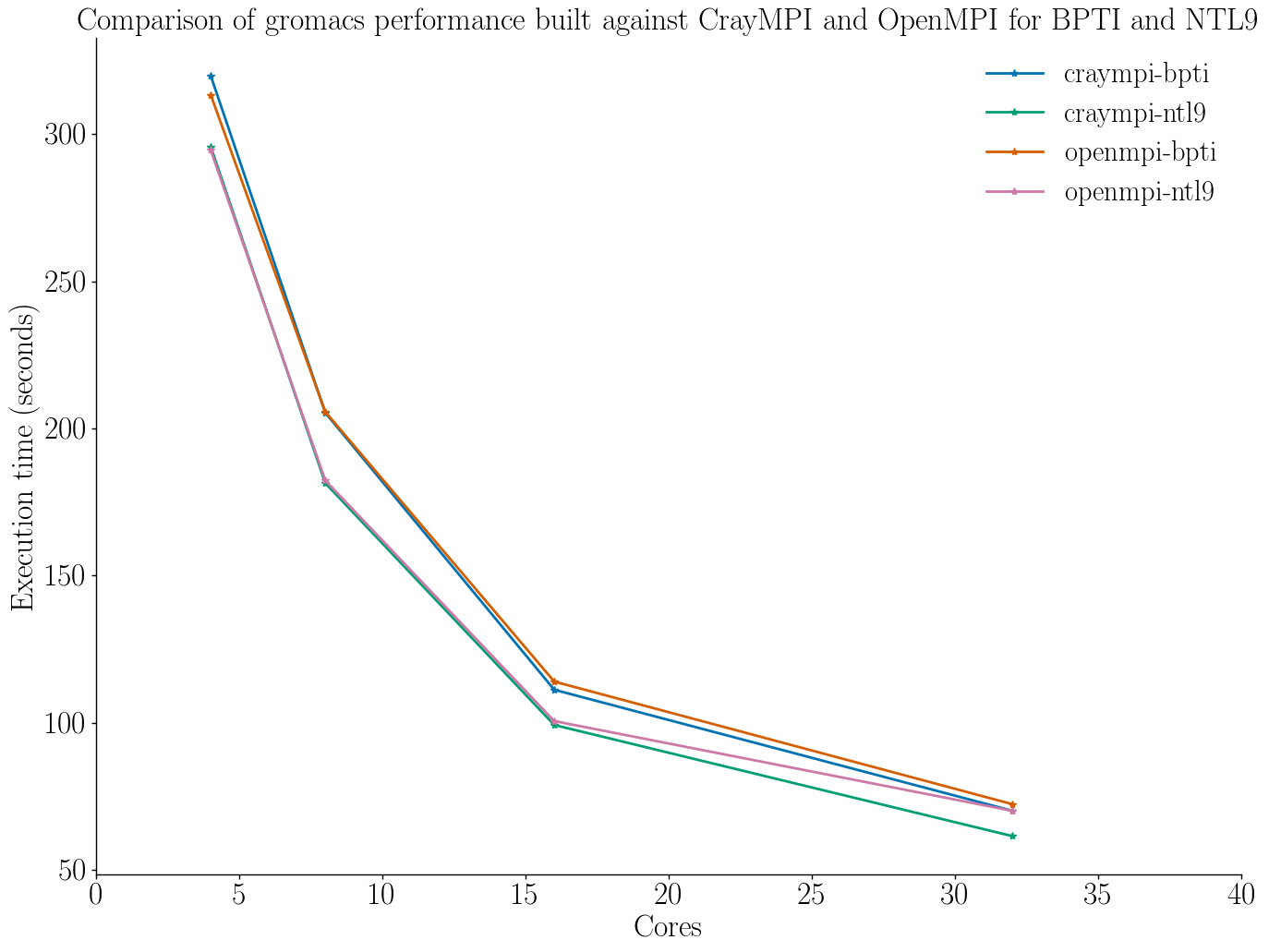}
    \caption{Gromacs scaling on \titan with OpenMP.}\label{fig:grom_omp}
\end{figure}

\begin{figure}[h!]
    \centering
    \includegraphics[width=0.9\linewidth]{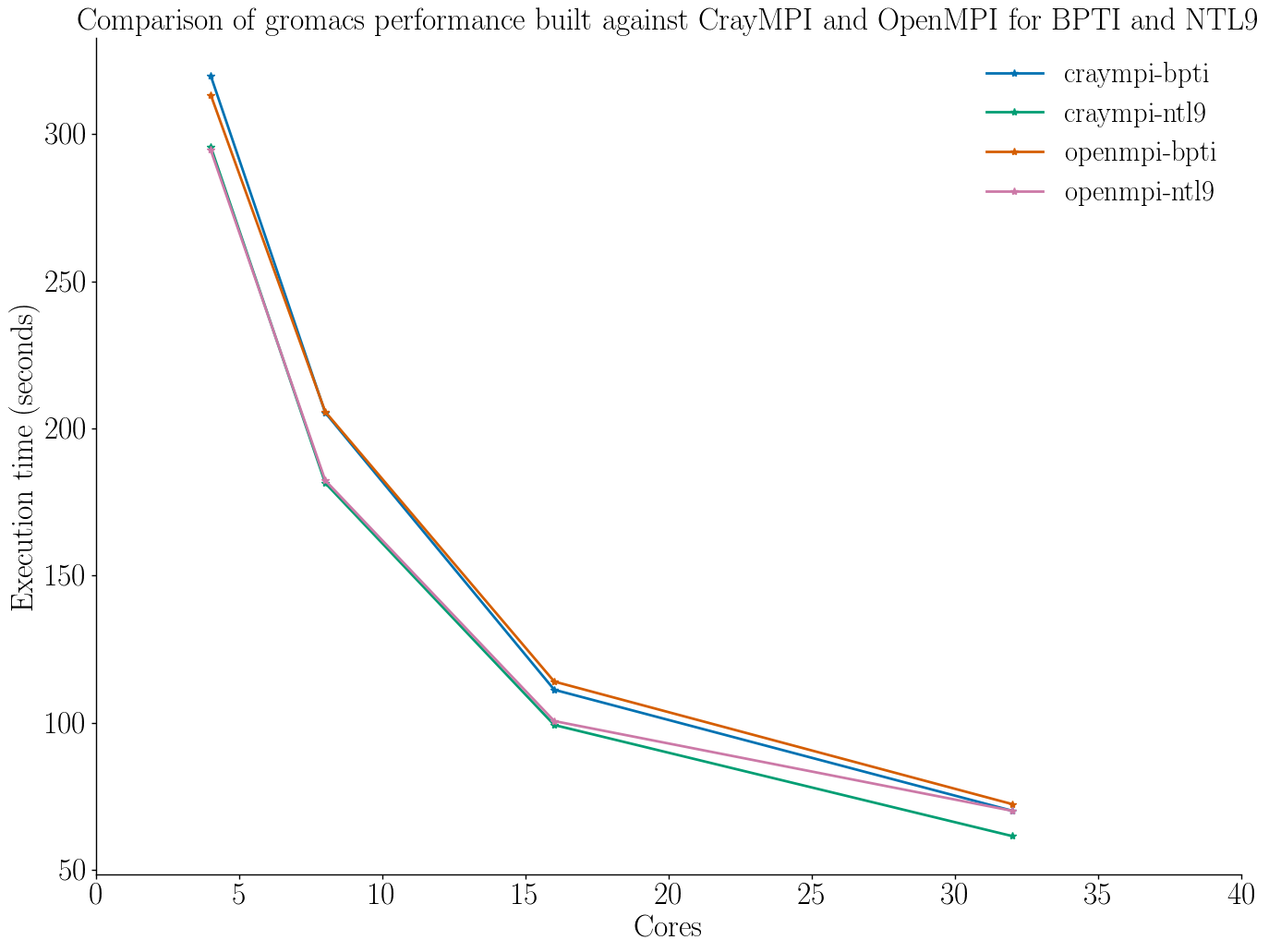}
    \caption{Gromacs scaling on \titan with OpenMPI.}\label{fig:grom_mpi}
\end{figure}

Figure~\ref{fig:omp} shows the emulation of a Gromacs workload when scaling
to a full node on \titan (16 cores) and \supermic (20 cores), respectively.
Values are shown for both, OpenMP and OpenMPI emulation.  We observe that
\supermic (Xeon, 2,8 GHz) executes the tasks faster than \titan (Opterons,
2.2 GHz). Interesting is that OpenMP outperforms OpenMPI on \titan, but we
observe the opposite on \supermic.  Also, \titan's performance is more
consistent (smaller error bars).  In both cases though, we find the scaling
behavior to be similar to the actual Gromacs application (see 
Figures~\ref{fig:grom_omp}, \ref{fig:grom_mpi}).

\subsubsection*{Experiment E.5: Emulating Variable I/O Granularity}

\begin{figure}[!h]
  \centering
  \includegraphics[width=0.85\linewidth]{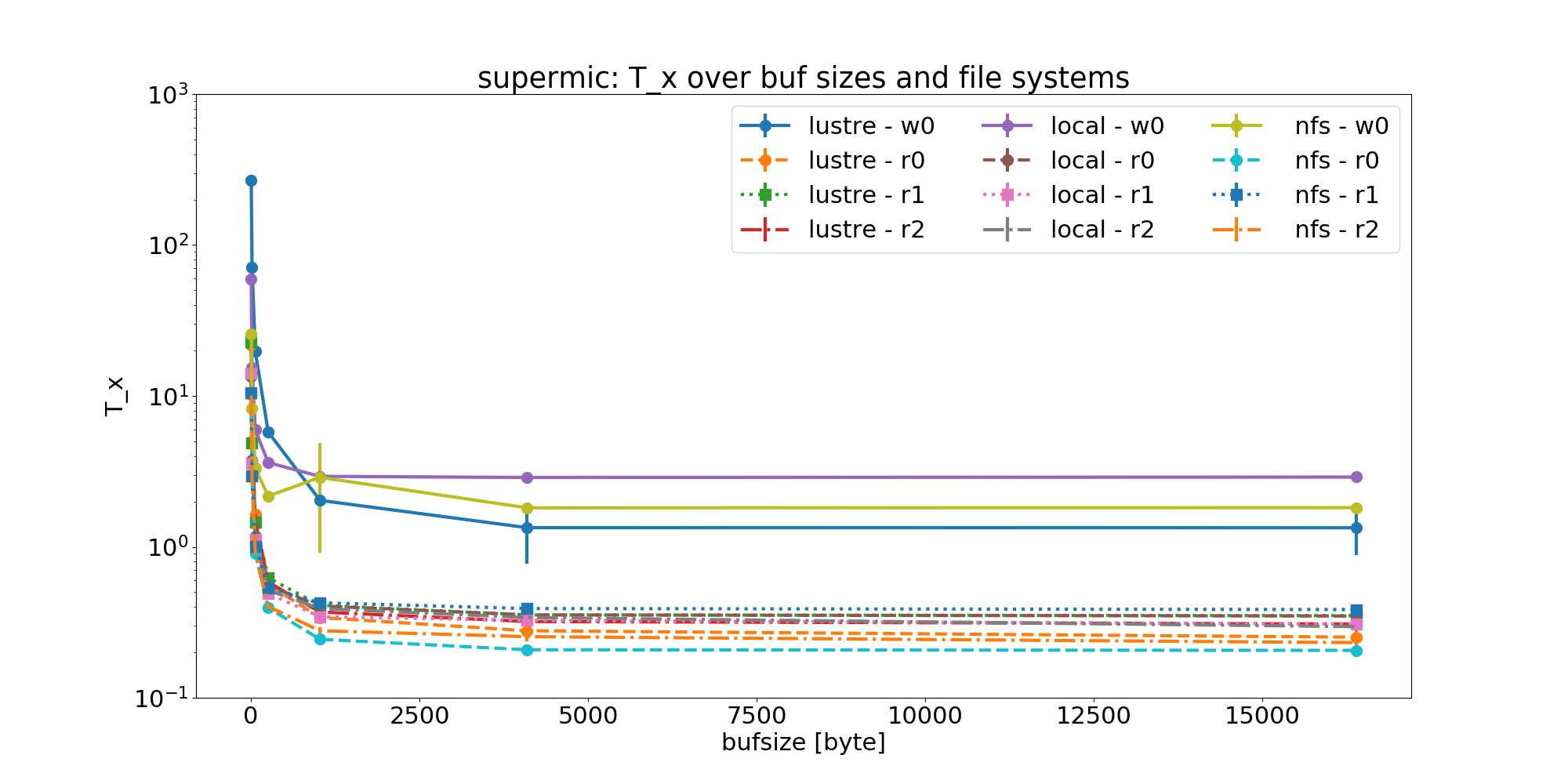}
  \includegraphics[width=0.85\linewidth]{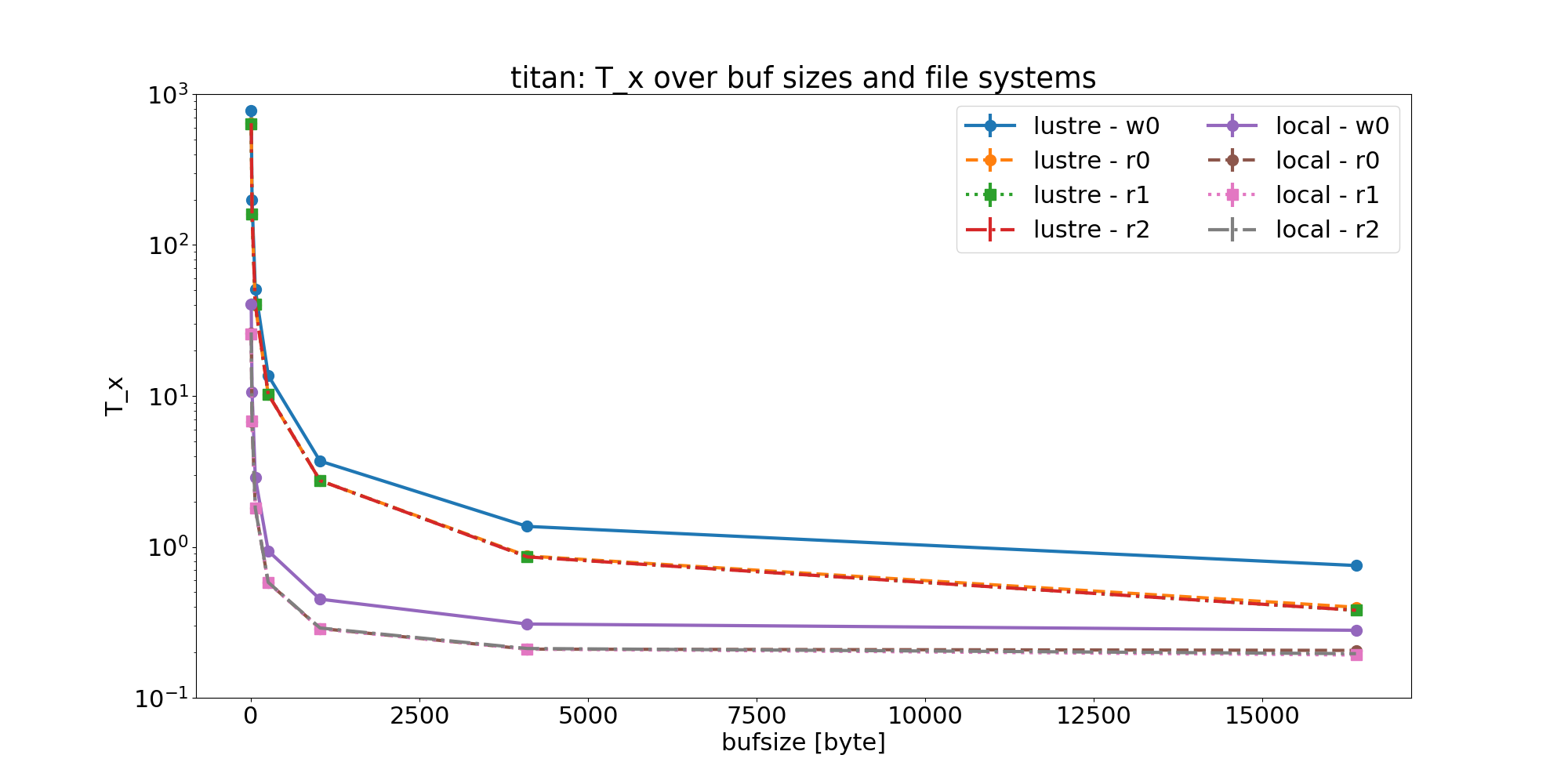}
  \caption{
    \textbf{I/O Emulation:} The emulation of application I/O is malleable in
    two dimensions; toward the resource filesystem to be used as target, and
    toward the I/O block size.  The resulting performance of a static and
    homogeneous set of I/O operations varies as expected: smaller block
    imply larger I/O cost; distributed file systems are an order of magnitude
    slower than local ones; Lustre is caching more aggressively than NSF, and
    performs about the same on \titan (top) and \supermic (bottom). 
    The local FS on \titan performs much better than the one on \supermic 
    though.\label{fig:io} 
  }
\end{figure}

\Synapse's is not yet able to profile application I/O granularity.
Nevertheless, \synapse \I{emulation} can be tuned in great detail: the I/O
can be emulated toward any available filesystem, any number of files, and any
combination of I/O granularity for those files.  The experiments shown in
Figure~\ref{fig:io} demonstrate those capabilities.

The experiments vary the application emulation in two dimensions: change of
the file system used for all I/O operations, and the granularity for all I/O
operations. The figures show that write operations are generally an order of
magnitude slower than read operations, which is owed to the difficulty of
providing cache consistency on write, specifically on shared file systems.
They further show that many small I/O operations are much slower than I/O for
large blocks of data, clearly showing the influence of the file system
latencies.  It is interesting to see that Lustre performs very similar for
both resources (\supermic and \titan), whereas local I/O performance differs
significantly.

We want to refrain from interpreting the results in too much detail, as the
intent is not to discuss the actual I/O performance of the underlying
filesystem (as interesting as that information though is for our own
research).  The value of the measured information is also limited, as the
file systems were only used for two neighboring nodes (which likely access
the same Lustre metadata service and I/O node).  The granularity is here
considered constant over the lifetime of the application, which is also
unrealistic, and emulating actual application I/O granularity will yield very
different results.  However, Synapse is well able to tune a representative
application workloads in different I/O dimensions, which expands its usage
scope to a range of use cases beyond those centered on computationally
dominated applications.

%
\section{Future Work}\label{sec:future}

\subsubsection*{Profiling Block-Level I/O Operations}

The performance of disk I/O operations depends heavily on the storage system
that is used, and on the granularity of the I/O requests toward that storage
system. \synapse currently captures neither of those, but we plan to use
|blktrace| to obtain those information. Currently, only a prototype watcher
plugin for |blktrace| exists.

\subsubsection*{Sampling Rate}

A high sampling rate has been shown to be able to capture application startup
more accurately, and is necessary to profile short-running jobs. At the same
time, a high sampling rate creates large profile traces, and despite the
limited measured overhead we want to be cautious of the side effects of
high-frequency sampling. We thus consider an adaptive scheme, starting with a
high sampling rate (10/sec), and after a few seconds, when we can expect to
have captured the application startup, decrease the rate. \synapse's codebase
does not assume a constant rate, and can be therefore extended to adaptively
adjust the sampling rate.

\subsubsection*{Profiling and Emulation of Networking, MPI}

The most significant and challenging improvement is profiling and emulation
of data sent over network connections.  This would require changes to our
current profiling approach as a sample-based inspection seems insufficient to
capture that information. We consider using |libc| call tracing for that
purpose. A similar route seems useful to support the profiling and emulation
of MPI and OpenMP applications.  A wide variety of MPI and OpenMP tracing
tools and libraries exists. We plan to investigate how these libraries can be
used by \synapse to profile the execution of an application.

%
\section{Related Work}\label{sec:related}

PAPI~\cite{papi2004} is widely used in the HPC community. PAPI's sample based
evaluation of hardware counters is conceptually similar to \synapse
profiling. The simpler version in \synapse is based on standard Linux system
utilities, motivated by the use of resources where PAPI was not available and
where we lacked permissions to install it. Also, using \texttt{perf} and
other Linux tools integrated better with some elements of \synapse profiling,
such as disk I/O or memory allocation, which are not covered by PAPI\@. There
will likely be convergence with PAPI for some of \synapse's profiling needs,
so as to make the \synapse profiling more portable and easier to
maintain.

We are aware of only a few efforts to combine non-intrusive application
profiling with application emulation. In~\cite{sodhi2004skeleton}, the
authors describe an approach to automatically derive application
characteristics. It focuses on tracing the application's communication calls,
including MPI calls. Other resource interactions are considered opaque and
measured as times, and are thus system dependent.  This approach works well
for communication bound applications.  The emulation represents a subset of
the application: application $T_x$ is extrapolated under the assumption that
the subset is representative.

In~\cite{skeleton-synapse} Katz et al. work on a complementary approach of
\I{Application Skeletons}.  Skeletons do not include any mechanisms for
automatic application profiling, and thus require the user to specify
resource consumptions manually.  The focus of Skeletons is primarily on the
representation of logical and data dependencies between individual
application components: Application Skeletons can be used to represent a DAG
of such components.  Ref.~\cite{skeleton-synapse} discusses how \synapse can
be used to complement Application Skeletons, in that it provides
configuration parameters at the level of individual DAG components.

A large body of work on the \I{simulation} of application execution exists,
which aims to predict application runtimes (and other metrics) based on certain
models of resources and runtime environments.  A notable example in our context
of distributed computing is SimGrid~\cite{simgrid}: it provides a number of
interfaces (including the native application code) to "execute" an application
on a simulated distributed environment.  It covers communication,
virtualization,  data storage, different levels of systems and application
scheduling, etc., and has been successfully used in a large number of projects
to predict application performance on a specific target platform.

SimGrid and comparable simulation approaches are fundamentally different from
\synapse in that they use a truthful model of the target resource to simulate
the application execution: the better that model, the better the prediction of
application runtime behavior.  \Synapse does not use such a model, but instead
uses the \I{actual} target resource to execute a representation of the
application.  \Synapse is also not predictive: it actually executes the
application representation and measures its behavior.

%
\section{Conclusions}\label{sec:conclusions}

\Synapse is capable of automatically capturing essential application
characteristics, and of configuring representative application emulation.
While the application used to validate \synapse (Gromacs) is representative
of many other applications used in computational science\amnote{Should we
support this statement?}\mtnote{It is very generic and encompassing so I do
not think we should add more}, it remains to be seen if this approach can
suitably extend toward other application domains, and specifically towards
scenarios including inter-process communication.

The profiling capability of \synapse has a low runtime overhead, and provides
stable, consistent results. It requires no human intervention, code
instrumentation, or exchange of libraries, and is fully transparent to the
application. However, the profiling capability of \synapse requires support
at the system level and is constrained to resources where |perf stat| and
other system tools can be executed by users. We believe this not to be an
issue in practice.\mtnote{Why?}

The emulation capabilities of \synapse provide a stable, tuneable, malleable
and relatively accurate representation of the application's behavior, within
constraints. When used on the same resource as where the application was
profiled, \Synapse provides high-fidelity emulation of the target
applications. When used on resource different from where the application is
profiled, the emulation still manages to capture the essential application
characteristics and important trends that determine its execution time
\(T_x\). 


The main contribution to emulation uncertainties arise from resource-specific
compile-time optimizations of the application code, which are not captured
when using application profiles on resources different than the one where the
application is profiled. However, \synapse allows users to write, build and
use their own kernels to improve the fidelity of their emulations. The
ability to provide their 
own kernels allows users to control the way in which system resources are
consumed during emulation, as well as the libraries and functions that are
used in the kernel. The ability to control how kernels are built allows users
to apply resource-specific compile-time optimizations that are similar to
those used on the original application.

Given the simplicity and low usage overhead, we believe \synapse provides
a useful contribution to the computational science community. In fact,
\synapse is alreadyd used as a proxy application for the three use cases
discussed in Section~2.

%
\paragraph*{Software Availability}\label{par:software}

\synapse is available as Open Source Software, under the LGPL license,
at~\cite{synapse_url}.  The experiments in this paper used version
|v0.10|.  All scripts and configurations, along with the raw data
sets and scripts for plotting, are available at~\cite{synapse_exp}.
Please refer to the |README.md| file for instructions on how to
reproduce the experiments.  Comments, feedback, and contributions to
the software are welcome.  A bugtracker (which can also be used for
feedback) is available at~\cite{synapse_tracker}.  When using this
software, please reference~\cite{synapse}.


%
\section*{Acknowledgements}

\footnotesize{This work is supported by NSF ``CAREER'' ACI-1253644, NSF
ACI-1440677 ``RADICAL-Cybertools'' and DOE Award DE-SC0008651. We acknowledge
access to computational facilities on XSEDE resources via TG-MCB090174. We
thank members of the RADICAL group, in particular Vivek Balasubramanian and
Mark Santcroos for testing. We also thank Daniel S. Katz (U. Chicago) for
useful discussions in the context of Skeletons and for improvements to the
paper.}

%

\bibliographystyle{elsarticle-num}
\bibliography{2016-pdsec,enmd-local}

\end{document}